\documentclass[twocolumn, tighten]{aastex63}

\usepackage[caption=false]{subfig}
\usepackage{xfrac}

\shorttitle{GM Aur CO Structures}
\shortauthors{Huang et al.}

\begin{document}

\title{Molecules with ALMA at Planet-forming Scales (MAPS) XIX. Spiral Arms, a Tail, and Diffuse Structures Traced by CO around the GM Aur Disk}

\correspondingauthor{Jane Huang}
\email{jnhuang@umich.edu}
\author[0000-0001-6947-6072]{Jane Huang}
\altaffiliation{NASA Hubble Fellowship Program Sagan Fellow}
\affiliation{Department of Astronomy, University of Michigan, 323 West Hall, 1085 S. University Avenue, Ann Arbor, MI 48109, USA}
\affiliation{Center for Astrophysics \textbar\ Harvard \& Smithsonian, 60 Garden St., Cambridge, MA 02138, USA}
\author[0000-0003-4179-6394]{Edwin A. Bergin} \affiliation{Department of Astronomy, University of Michigan, 323 West Hall, 1085 S. University Avenue, Ann Arbor, MI 48109, USA}
\author[0000-0001-8798-1347]{Karin I. \"Oberg} \affiliation{Center for Astrophysics \textbar\ Harvard \& Smithsonian, 60 Garden St., Cambridge, MA 02138, USA}
\author[0000-0003-2253-2270]{Sean M. Andrews} \affiliation{Center for Astrophysics \textbar\ Harvard \& Smithsonian, 60 Garden St., Cambridge, MA 02138, USA}
\author[0000-0003-1534-5186]{Richard Teague}\affiliation{Center for Astrophysics \textbar\ Harvard \& Smithsonian, 60 Garden St., Cambridge, MA 02138, USA}
\author[0000-0003-1413-1776]{Charles J. Law}\affiliation{Center for Astrophysics \textbar\ Harvard \& Smithsonian, 60 Garden St., Cambridge, MA 02138, USA}
\author[0000-0002-6221-5360]{Paul Kalas}
\affiliation{Department of Astronomy, 501 Campbell Hall, University of California, Berkeley, CA 94720-3411, USA}
\affiliation{SETI Institute, Carl Sagan Center, 189 Bernardo Ave.,  Mountain View, CA 94043, USA}
\affiliation{Institute of Astrophysics, FORTH, GR-71110 Heraklion, Greece}
\author[0000-0003-3283-6884]{Yuri Aikawa}
\affiliation{Department of Astronomy, Graduate School of Science, The University of Tokyo, 7-3-1 Hongo, Bunkyo-ku, Tokyo 113-0033, Japan}
\author[0000-0001-7258-770X]{Jaehan Bae}
\altaffiliation{NASA Hubble Fellowship Program Sagan Fellow}
\affil{Department of Astronomy, University of Florida, Gainesville, FL 32611, USA}
\affil{Earth and Planets Laboratory, Carnegie Institution for Science, 5241 Broad Branch Road NW, Washington, DC 20015, USA}
\author[0000-0002-8716-0482]{Jennifer B. Bergner}\altaffiliation{NASA Hubble Fellowship Program Sagan Fellow} \affiliation{Department of Geophysical Sciences, University of Chicago, Chicago, IL 60637, USA}

\author[0000-0003-2014-2121]{Alice S. Booth} 
\affiliation{Leiden Observatory, Leiden University, 2300 RA Leiden, the Netherlands}
\affil{School of Physics \& Astronomy, University of Leeds, Leeds LS2 9JT, UK}
\author[0000-0003-4001-3589]{Arthur D. Bosman} \affiliation{Department of Astronomy, University of Michigan, 323 West Hall, 1085 S. University Avenue, Ann Arbor, MI 48109, USA}
\author[0000-0002-0150-0125]{Jenny K. Calahan} 
\affiliation{Department of Astronomy, University of Michigan, 323 West Hall, 1085 S. University Avenue, Ann Arbor, MI 48109, USA}
\author[0000-0002-2700-9676]{Gianni Cataldi}
\affil{National Astronomical Observatory of Japan, Osawa 2-21-1, Mitaka, Tokyo 181-8588, Japan}
\affil{Department of Astronomy, Graduate School of Science, The University of Tokyo, 7-3-1 Hongo, Bunkyo-ku, Tokyo 113-0033, Japan}
\author[0000-0003-2076-8001]{L. Ilsedore Cleeves}
\affiliation{University of Virginia, 530 McCormick Rd, Charlottesville, VA 22904, USA}
\author[0000-0002-1483-8811]{Ian Czekala}
\altaffiliation{NASA Hubble Fellowship Program Sagan Fellow}
\affiliation{Department of Astronomy and Astrophysics, 525 Davey Laboratory, The Pennsylvania State University, University Park, PA 16802, USA}
\affiliation{Center for Exoplanets and Habitable Worlds, 525 Davey Laboratory, The Pennsylvania State University, University Park, PA 16802, USA}
\affiliation{Center for Astrostatistics, 525 Davey Laboratory, The Pennsylvania State University, University Park, PA 16802, USA}
\affiliation{Institute for Computational \& Data Sciences, The Pennsylvania State University, University Park, PA 16802, USA}
\affiliation{Department of Astronomy, 501 Campbell Hall, University of California, Berkeley, CA 94720-3411, USA}
\author[0000-0003-1008-1142]{John~D.~Ilee} \affil{School of Physics \& Astronomy, University of Leeds, Leeds LS2 9JT, UK}
\author[0000-0003-1837-3772]{Romane Le Gal}
\affiliation{Univ. Grenoble Alpes, CNRS, IPAG, F-38000 Grenoble, France}
\affiliation{IRAM, 300 rue de la piscine, F-38406 Saint-Martin d'H\`{e}res, France}
\affiliation{IRAP, Universit\'e de Toulouse, CNRS, CNES, UT3, Toulouse, France}
\affiliation{Center for Astrophysics \textbar\ Harvard \& Smithsonian, 60 Garden St., Cambridge, MA 02138, USA}
\author[0000-0003-4784-3040]{Viviana V. Guzm\'{a}n}
\affiliation{Instituto de Astrof\'isica, Pontificia Universidad Cat\'olica de Chile, Av. Vicu\~na Mackenna 4860, 7820436 Macul, Santiago, Chile}
\author[0000-0002-7607-719X]{Feng Long}
\affiliation{Center for Astrophysics \textbar\ Harvard \& Smithsonian, 60 Garden St., Cambridge, MA 02138, USA}
\author[0000-0002-8932-1219]{Ryan A. Loomis}\affiliation{National Radio Astronomy Observatory, 520 Edgemont Rd., Charlottesville, VA 22903, USA}
\author[0000-0002-1637-7393]{Fran\c cois M\'enard}\affiliation{Univ. Grenoble Alpes, CNRS, IPAG, F-38000 Grenoble, France}
\author[0000-0002-7058-7682]{Hideko Nomura}\affil{National Astronomical Observatory of Japan, Osawa 2-21-1, Mitaka, Tokyo 181-8588, Japan}
\author[0000-0001-8642-1786]{Chunhua Qi} \affiliation{Center for Astrophysics \textbar\ Harvard \& Smithsonian, 60 Garden St., Cambridge, MA 02138, USA}
\author[0000-0002-6429-9457]{Kamber R. Schwarz} \altaffiliation{NASA Hubble Fellowship Program Sagan Fellow}
\affiliation{Lunar and Planetary Laboratory, University of Arizona, 1629 E. University Blvd, Tucson, AZ 85721, USA}
\author[0000-0002-6034-2892]{Takashi Tsukagoshi} \affil{National Astronomical Observatory of Japan, Osawa 2-21-1, Mitaka, Tokyo 181-8588, Japan}
\author[0000-0002-2555-9869]{Merel L. R. van 't Hoff}
\affiliation{Department of Astronomy, University of Michigan, 323 West Hall, 1085 S. University Avenue, Ann Arbor, MI 48109, USA}
\author[0000-0001-6078-786X]{Catherine Walsh}\affiliation{School of Physics \& Astronomy, University of Leeds, Leeds LS2 9JT, UK}
\author[0000-0003-1526-7587]{David J. Wilner}\affiliation{Center for Astrophysics \textbar\ Harvard \& Smithsonian, 60 Garden St., Cambridge, MA 02138, USA}
\author[0000-0003-4099-6941]{Yoshihide Yamato} \affiliation{Department of Astronomy, Graduate School of Science, The University of Tokyo, 7-3-1 Hongo, Bunkyo-ku, Tokyo 113-0033, Japan}
\author[0000-0002-0661-7517]{Ke Zhang}
\altaffiliation{NASA Hubble Fellow}
\affiliation{Department of Astronomy, University of Wisconsin-Madison, 
475 N Charter St, Madison, WI 53706}
\affiliation{Department of Astronomy, University of Michigan, 
323 West Hall, 1085 S. University Avenue, 
Ann Arbor, MI 48109, USA}

\begin{abstract}
The concentric gaps and rings commonly observed in protoplanetary disks in millimeter continuum emission have lent the impression that planet formation generally proceeds within orderly, isolated systems. While deep observations of spatially resolved molecular emission have been comparatively limited, they are increasingly suggesting that some disks interact with their surroundings while planet formation is underway. We present an analysis of complex features identified around GM Aur in $^{12}$CO $J=2-1$ images at a spatial resolution of $\sim40$ au. In addition to a Keplerian disk extending to a radius of $\sim550$ au, the CO emission traces flocculent spiral arms out to radii of $\sim$1200 au, a tail extending $\sim1800$ au southwest of GM Aur, and diffuse structures extending from the north side of the disk up to radii of $\sim1900$ au. The diffuse structures coincide with a ``dust ribbon" previously identified in scattered light. The large-scale asymmetric gas features present a striking contrast with the mostly axisymmetric, multi-ringed millimeter continuum tracing the pebble disk. We hypothesize that GM Aur's complex gas structures result from late infall of remnant envelope or cloud material onto the disk. The morphological similarities to the SU Aur and AB Aur systems, which are also located in the L1517 cloud, provide additional support to a scenario in which interactions with the environment are playing a role in regulating the distribution and transport of material in all three of these Class II disk systems. This paper is part of the MAPS special issue of the Astrophysical Journal Supplement.   
\end{abstract}

\keywords{protoplanetary disks---ISM: molecules---stars: individual (GM Aur)}

\section{Introduction} \label{sec:intro}
Observing the physical and chemical conditions in protoplanetary disks is key for elucidating how planets form. Carbon monoxide (CO) has long been one of the most important probes of disk structure and evolution because of its high abundance in disks and the existence of readily detectable rotational transitions at millimeter wavelengths. Early interferometric CO detections toward pre-main sequence stars established the presence of gas-rich, Keplerian disks \citep[e.g.,][]{1987ApJ...323..294S, 1993Icar..106....2K}. Disks with detections of bright, extended Keplerian CO emission naturally became popular targets for chemical studies at millimeter wavelengths, initially with single-dish telescopes \citep[e.g.,][]{1997AA...317L..55D, 2004AA...425..955T} and then with interferometers \citep[e.g.][]{2003ApJ...597..986Q, 2010ApJ...720..480O}. 

Five such disks (IM Lup, GM Aur, AS 209, HD 163296, and MWC 480) constitute the sample for Molecules with ALMA at Planet-forming Scales (MAPS), an ALMA Cycle 6 Large Program that conducted the first high angular resolution millimeter wavelength survey of protoplanetary disk chemistry \citep{2021arXiv210906268O}. These sources were selected because of the rich and varied substructures that have previously been imaged with the Atacama Large Millimeter/Submillimeter Array (ALMA) in a variety of molecules \citep[e.g.,][]{2017ApJ...835..231H, 2019ApJ...876...25B, 2020ApJ...890..142P} as well as in millimeter continuum emission \citep[e.g.,][]{2018ApJ...869L..41A,2018ApJ...869L..42H, 2018ApJ...869...17L, 2020ApJ...891...48H}. 

The distribution and transportation of gas are key inputs for models of disk evolution and planet formation. Simulations of the formation of dust gaps and rings via planet-disk interactions typically adopt simple, axisymmetric gas surface density profiles, with no exchange of material with their environments \citep[e.g.,][]{2015MNRAS.453L..73D, 2017ApJ...850..201B,  2018ApJ...869L..47Z}. While deep molecular line observations support this treatment for some widely modelled systems, including most of the MAPS targets, recent observations of systems such as HL Tau, RU Lup, and BHB1 have shown that concentric gap and ring dust structures can co-exist with large-scale, non-Keplerian stream or spiral structures associated with the disk \citep[e.g.,][]{2015ApJ...808L...3A, 2017AA...608A.134Y, 2018ApJ...869L..41A, 2019ApJ...880...69Y, 2020ApJ...898..140H, 2020ApJ...904L...6A}. However, since most ALMA CO disk observations have been perfunctory, with integration times on the order of a couple minutes \citep[e.g.,][]{2016ApJ...827..142B, 2018ApJ...859...21A}, the range and prevalence of possible gas structures within and around disks are not yet well characterized.

In this article, we analyze complex CO features identified in MAPS observations of the disk around GM Aur (ICRS 04:55:10.981, +30:21:59.376), a K5.5 T Tauri star at a distance of 159 pc in the Taurus-Auriga star-forming region \citep[e.g.,][]{2010ApJ...717..441E, 2018AA...616A...1G, 2018AJ....156...58B}. As one of the first disks to be identified as ``transitional'' (i.e., with a central dust cavity) and to be spatially resolved in CO emission, it has been subject to extensive observations from centimeter to X-ray wavelengths in order to probe the disk structure, chemistry, accretion behavior, and signatures of planet-disk interactions  \citep[e.g.,][]{1989AJ.....97.1451S, 1993Icar..106....2K, 2004ApJ...614L.133B, 2005ApJ...630L.185C, 2009ApJ...698..131H, 2019ApJ...877L..34E}. Much like most other disks that have now been imaged at moderate-to-high resolution, the GM Aur disk's millimeter continuum displays a series of nearly axisymmetric gaps and rings, \citep[e.g.,][]{2020ApJ...891...48H}. Meanwhile, low-resolution ($\sim2''$) $^{12}$CO observations have traced large-scale asymmetric structures deviating from the expected Keplerian velocity field of the disk \citep{2008AA...490L..15D, 2009ApJ...698..131H}. The MAPS program improves upon previously published GM Aur $^{12}$CO observations by an order of magnitude in angular resolution, revealing a tremendously intricate system with flocculent spiral arms, a tail, and diffuse structures. This large-scale complexity sets GM Aur apart from the other MAPS disks, which exhibit nearly axisymmetric CO emission \citep{2021arXiv210906210L}. The GM Aur observations and data reduction are presented in Section \ref{sec:observations}, the molecular features are described in Section \ref{sec:results}, their possible origins are discussed in Section \ref{sec:discussion}, and the conclusions are summarized in Section \ref{sec:conclusions}.

\section{Observations and Data Reduction \label{sec:observations}}
\subsection{ALMA observations}
A full description of the observational setup and self-calibration procedure for the MAPS Large Program is provided in \citet{2021arXiv210906268O}.\footnote{Self-calibration and imaging scripts, as well as imaging products, are available at \url{http://alma-maps.info}} We briefly summarize the key aspects. GM Aur was observed in the spectral setting containing $^{12}$CO $J=2-1$ over the course of eight execution blocks shared with MWC 480. Three were taken in the C43-4 configuration (baselines ranging from 14 to 1397 m) in 2018 October and November for a combined on-source integration time of one hour. The other five were taken in the C43-7 configuration (baselines ranging from 40 to 3638 m) in 2019 August, for a combined on-source integration time of two hours. Following self-calibration, a set of fiducial images was produced at a spatial resolution of $0\farcs15$ and spectral resolution of 0.2 km s$^{-1}$ for all Band 6 lines covered in the survey \citep{2021arXiv210906188C}. The fiducial imaging procedure aimed to facilitate high-resolution studies of the radial and vertical distributions of molecular emission. Analysis of the structure and chemistry of GM Aur's Keplerian disk region as traced by CO isotopologues is presented in \citet{2021arXiv210906210L, 2021arXiv210906217L, 2021arXiv210906233Z,2021arXiv210906223B, 2021arXiv210906228S}. 

However, GM Aur's $^{12}$CO emission is visible in some channels up to $\sim12''$ from the phase center. The ALMA Technical Handbook\footnote{\url{https://almascience.nrao.edu/documents-and-tools/cycle7/alma-technical-handbook/view}} suggests two definitions for maximum recoverable scale (MRS), one calculated using the shortest baseline $L_{\text{min}}$ ($\theta_{\text{MRS}}\approx 0.6\lambda_{\text{obs}}/L_{\text{min}}$) and the more conservative one using the 5th percentile of the baseline lengths ($\theta_{\text{MRS}}\approx 0.983\lambda_{\text{obs}}/L_{\text{5}}$). Given these definitions, the estimated MRS of GM Aur's CO emission ranges from $\sim$2$''$ to 11$''$. To improve sensitivity to large-scale emission, we re-imaged $^{12}$CO $J=2-1$ with a Gaussian $uv$ taper and a robust value of 1.0, resulting in a synthesized beam of $0\farcs28\times0\farcs23$ (-14\fdg4). Unless otherwise specified, figures in this article are based on this version of the $^{12}$CO image, which we will subsequently refer to as the MAPS XIX version. The re-imaging was performed with \texttt{CASA 5.6.1} \citep{2007ASPC..376..127M}. We applied the multi-scale CLEAN algorithm \citep{2008ISTSP...2..793C} with scales of $[0, 0\farcs4, 1'', 2'']$. Since $^{12}$CO was also re-observed at higher spectral resolution than the other lines, we re-imaged $^{12}$CO with channel widths of 0.1 km s$^{-1}$ for better recovery of the kinematic details. Due to the  irregular emission morphology, we used \texttt{CASA's} auto-multithresh algorithm \citep{2020PASP..132b4505K} to draw the CLEAN mask. The auto-multithresh algorithm searches the cube for significant emission, beginning with a relatively conservative mask and then expanding to encompass more emission during subsequent major cycles. The mask was initialized with full coverage of the primary beam from $5.2-6.4$ km s$^{-1}$, where the emission is the broadest, because auto-multithresh algorithm does not readily mask diffuse emission. After some experimentation, the following auto-multithresh parameters were selected: \texttt{sidelobethreshold=3.0}, \texttt{noisethreshold=4.0}, \texttt{lownoisethreshold=1.5}, and \texttt{minbeamfrac=0.3}. The CLEAN threshold was set to 5 mJy, corresponding to $\sim3\times$ the rms of line-free channels in the dirty image.  

As noted in \citet{1995AJ....110.2037J}, the default procedure at the end of CLEAN is to add the residuals (in units of Jy per dirty beam) to the CLEAN model convolved with the synthesized beam (in units of Jy per clean beam) under the assumption that these units are roughly equivalent. However, this assumption does not hold when the dirty beam deviates significantly from the Gaussian shape of the synthesized beam, as is the case for the MAPS dataset. Throughout the MAPS paper series, this unit mismatch is referred to as the ``JvM effect.'' To counteract this issue, the MAPS Large Program applied a ``JvM correction'' to all image products, where the residuals are scaled by a factor of $\epsilon = (\text{Clean beam volume}/\text{Dirty beam volume})$ before being added to the convolved CLEAN model \citep{2021arXiv210906188C}. As noted in \citet{2021arXiv210906188C}, lowering the CLEAN threshold does not obviate the need for the JvM correction because CLEAN would then tend erroneously to incorporate more noise into the CLEAN model, and also because the units of the residuals would still be inconsistent with those of the convolved CLEAN model. We refer the reader to \citet{2021arXiv210906188C} for full details on the MAPS imaging workflow and the impact of the JvM effect and correction on the MAPS datasets. For the re-imaged GM Aur $^{12}$CO observations presented in this work, $\epsilon = 0.46$. A primary beam correction was applied to the resulting image using \texttt{impbcor} in \texttt{CASA}. The resulting rms of the image cube, measured within a circle with a radius of $10''$ in line-free channels, is 1 mJy beam$^{-1}$. 

The fiducial images for $^{13}$CO $J=2-1$, which was observed simultaneously with $^{12}$CO $J=2-1$, exhibited hints of residual structure outside its Keplerian CLEAN mask. While we used the auto-multithresh algorithm for $^{12}$CO because its emission is bright and irregular, we chose to use a simple circular mask with a diameter of $11''$ to re-image $^{13}$CO in order to check that noise fluctuations were not being amplified by selective masking. The CLEAN threshold was set to 4 mJy, corresponding to $\sim3\times$ the rms of line-free channels in the dirty image. To increase sensitivity, a Gaussian $uv$ taper and robust value of 1.0 were applied, resulting in a synthesized beam of $0\farcs39\times0\farcs35$ (-4\fdg4). Since the $^{13}$CO spectral windows had coarser spectral resolution than $^{12}$CO, the $^{13}$CO image cube was produced with a channel spacing of 0.2 km s$^{-1}$. In other respects, the $^{13}$CO re-imaging followed the same procedure as the $^{12}$CO re-imaging. The resulting rms of the image cube, following residual scaling corrections with $\epsilon=0.60$, is 1 mJy beam$^{-1}$. 

\subsection{Archival Hubble Space Telescope observations}

For comparison with the ALMA data, Wide Field Planetary Camera 2 (WFPC2) observations of GM Aur from program HST-GTO/WF2-6223 (PI: Trauger) were retrieved from the Hubble Space Telescope (HST) archive (now the Mikulski Archive for Space Telescopes). Two exposures lasting 200 seconds each (datasets U2RD0406T and U2RD0409T) were taken on 1995 July 29 in the F675W filter (central wavelength: 6717 \AA) with the PC1 camera, which has a pixel scale of 0\farcs0455. The data available on the archive were processed with the standard OPUS pipeline \citep{1999ASPC..172..203S}. PSF subtraction was performed on GM Aur using the reference star HD 283572, which was observed as part of the same program. An independent reduction of these data was previously presented in \citet{1997IAUS..182..355S}. While HST images of GM Aur have been published at other wavelengths, we use the WFPC2 F675W image as the basis for comparison because the WFPC2 images presented in \citet{2016ApJ...829...65H} had shorter exposure times, and the field of view of the NICMOS coronagraphic images presented in \citet{2003AJ....125.1467S} is more limited. 

\begin{figure*}
\begin{center}
\includegraphics{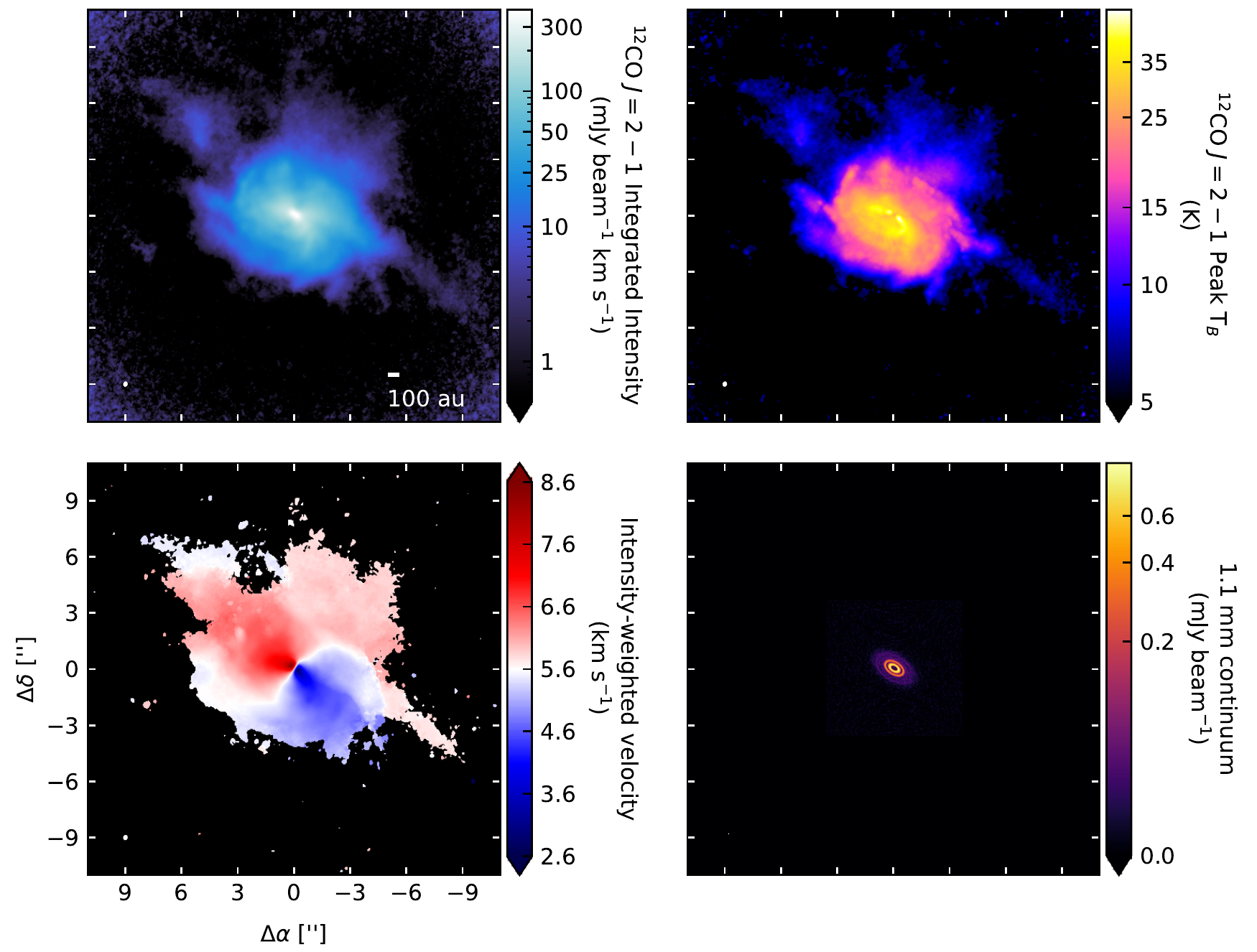}
\end{center}
\caption{Moment maps of $^{12}$CO $J=2-1$ toward GM Aur and a comparison with the millimeter continuum. Top left: An integrated intensity map of $^{12}$CO $J=2-1$ emission towards GM Aur. A logarithmic color stretch is used to highlight faint structures. The scale of the synthesized beam, $0\farcs28\times0\farcs23$ (-14\fdg4), is represented by the white ellipse in the lower left corner. Top right: A peak brightness temperature map of $^{12}$CO $J=2-1$. A logarithmic color stretch is used to highlight faint structures. Bottom left: The $^{12}$CO intensity-weighted velocity map. Bottom right: The high-resolution 1.1 mm GM Aur continuum image from \citet{2020ApJ...891...48H}. An arcsinh color stretch is used to highlight faint structures. The scale of the synthesized beam, 45 mas $\times$ 25 mas (2\fdg 2), is represented by the white ellipse in the lower left corner. All images are shown on the same scale to emphasize the different spatial extents of the molecular and dust emission. North is up and east is to the left. \label{fig:COoverview}}
\end{figure*}

\section{Results}\label{sec:results}
\subsection{Complex structures traced by $^{12}$CO}

Figure \ref{fig:COoverview} shows integrated intensity, intensity-weighted velocity, and peak brightness temperature maps of $^{12}$CO $J=2-1$ emission towards GM Aur, as well as the high resolution 1.1 mm continuum image from \citet{2020ApJ...891...48H} for comparison. The integrated intensity map includes channels with LSRK velocities from $-1.6$ to 12.5 km s$^{-1}$, based on the velocity range where disk emission is detected above the $3\sigma$ level. Because emission at any given position spans a much narrower velocity range than the overall velocity range encompassed by the integrated intensity map (see the $^{12}$CO $J=2-1$ channel maps in Appendix \ref{sec:chanmaps}), only pixels in individual channels above the $3\sigma$ level are included in the integrated intensity map in order to reduce noise contributions from signal-free regions. The fiducial integrated intensity maps presented in \citet{2021arXiv210906210L} use modified Keplerian masks, but this approach is not suitable for GM Aur's $^{12}$CO emission due to the extended, non-Keplerian structures. 

To produce a peak brightness temperature map, we imaged the $^{12}$CO line from visibilities that have not been continuum subtracted in order to avoid artificially reducing the intensities in regions where continuum emission is absorbed by optically thick $^{12}$CO \citep[e.g.,][]{2018ApJ...853..113W}. While the large-scale features of interest in this work lie outside the millimeter continuum and are therefore unaffected by continuum subtraction, we elect to make peak brightness temperature maps without continuum subtraction in order to obtain results consistent with \citet{2021arXiv210906217L}. We otherwise follow the imaging steps described in Section \ref{sec:observations}. Similar image cubes were also produced with the channel gridding shifted by $+0.03$ and $-0.03$ km s$^{-1}$. A map is generated for each cube by calculating the peak intensity as a function of velocity at each pixel. Emission below the $5\sigma$ level is masked. The three maps were then median-stacked to mitigate channelization artifacts \citep[e.g.,][]{2014ApJ...785L..12C}. The median peak intensity map was then converted to brightness temperatures using the Planck equation (i.e., not the Rayleigh-Jeans approximation). The intensity-weighted velocity map was created with the same velocity range as the other two maps, with pixels below the $5\sigma$ level masked. Higher masking thresholds were used for the peak intensity and intensity-weighted velocity maps compared to the integrated intensity map because the former are more sensitive to noisy outliers.  

The $^{12}$CO emission is broad and asymmetric, extending up to a distance of $\sim12''$ ($\sim1900$) au from the star. Several arm-like structures emerge from the edge of the disk, more prominently in the peak brightness temperature map compared to the integrated intensity map. A long ``tail'' extends from the southwest side of the disk. Diffuse emission surrounds the northern side of the disk. Spatial filtering afflicts the appearance of the largest-scale $^{12}$CO emission by creating negative CLEAN ``bowls'' in channels near the systemic velocity ($5.4-6.0$ km s$^{-1}$). However, the appearance of the extended structures in the central channels is broadly consistent with the emission morphology observed with the Submillimeter Array's more compact antenna configurations \citep{2009ApJ...698..131H}. The irregularity and large extent of the $^{12}$CO emission presents a stark contrast with the continuum emission, which is comparatively compact (extending up to a radius of $\sim$ 250 au) and nearly axisymmetric.

The $^{12}$CO velocity map traces the Keplerian disk out to a radius of $\sim550$ au, with the northeast side redshifted and the southwest side blueshifted relative to the systemic velocity of 5.61 km s$^{-1}$ \citep{2020ApJ...891...48H}. The extended emission exhibits more complex kinematic behavior\textemdash the southwest tail and diffuse northwest emission are slightly redshifted relative to the systemic velocity, the diffuse emission northeast of the disk becomes more blueshifted from south to north, and the southeastern edge of the $^{12}$CO emission is blueshifted relative to the Keplerian disk. The following sections examine the complex CO emission structures in more detail.

\subsubsection{The spiral arms}

\begin{figure*}
\begin{center}
\includegraphics{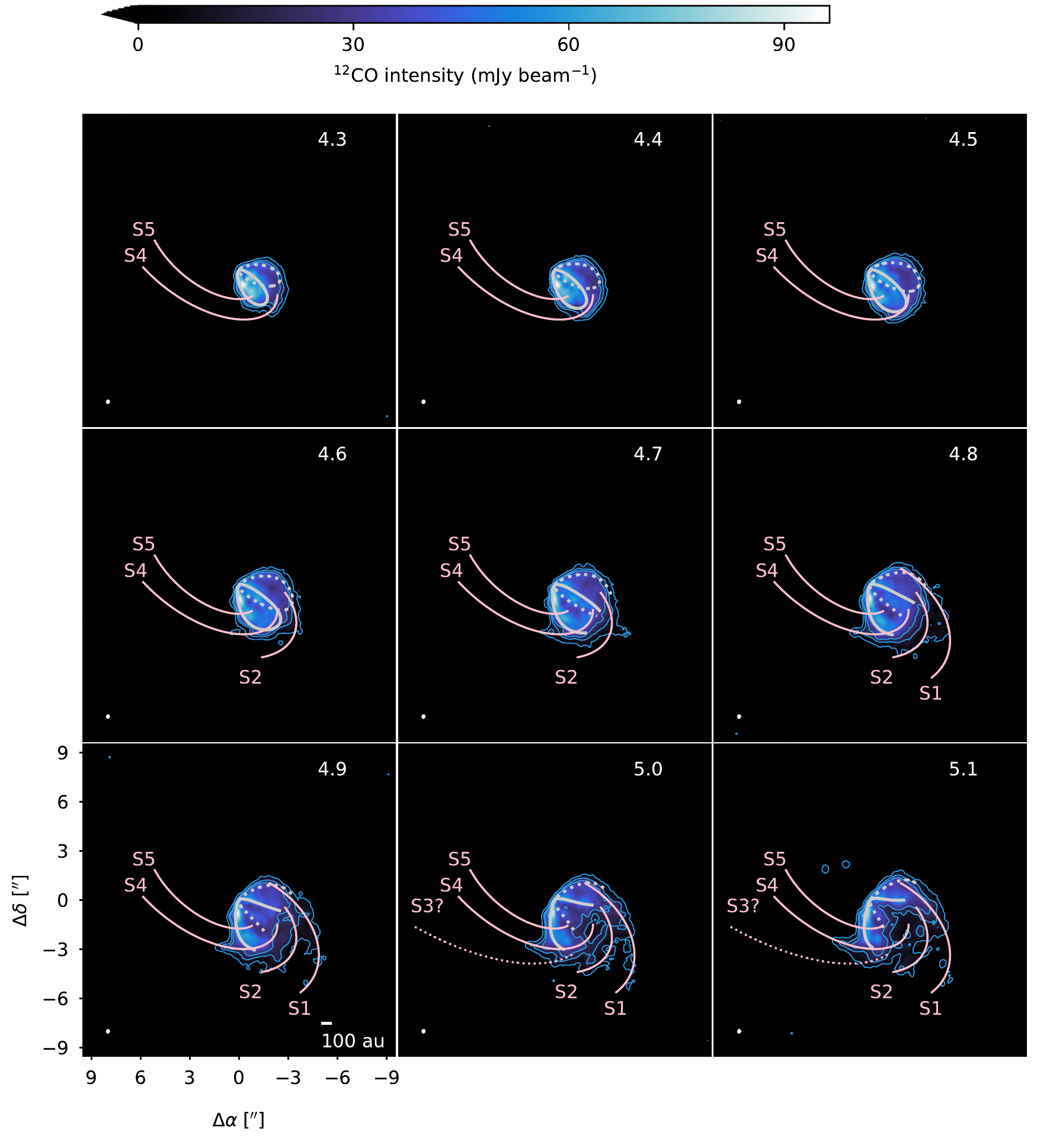}
\end{center}
\caption{$^{12}$CO channel maps toward GM Aur with pink curves overlaid on the gas spiral structures. The dotted curve for S3 denotes less certainty in its identification. The light blue contours show the 5, 15, and 25$\sigma$ emission levels, where $\sigma=1$ mJy beam$^{-1}$. The isovelocity contours of the Keplerian disk are drawn in light gray, with the solid curves tracing the front of the disk and the dotted curves tracing the back side. The upper right corner of each panel is labeled with the LSRK velocity (km s$^{-1}$). The scale of the synthesized beam, $0\farcs28\times0\farcs23$ (-14\fdg4), is represented by the white ellipse in the lower left corner of each panel. North is up and east is to the left. \label{fig:spiralchanmapspt1}}
\end{figure*}

\begin{figure*}
\begin{center}
\ContinuedFloat
\includegraphics{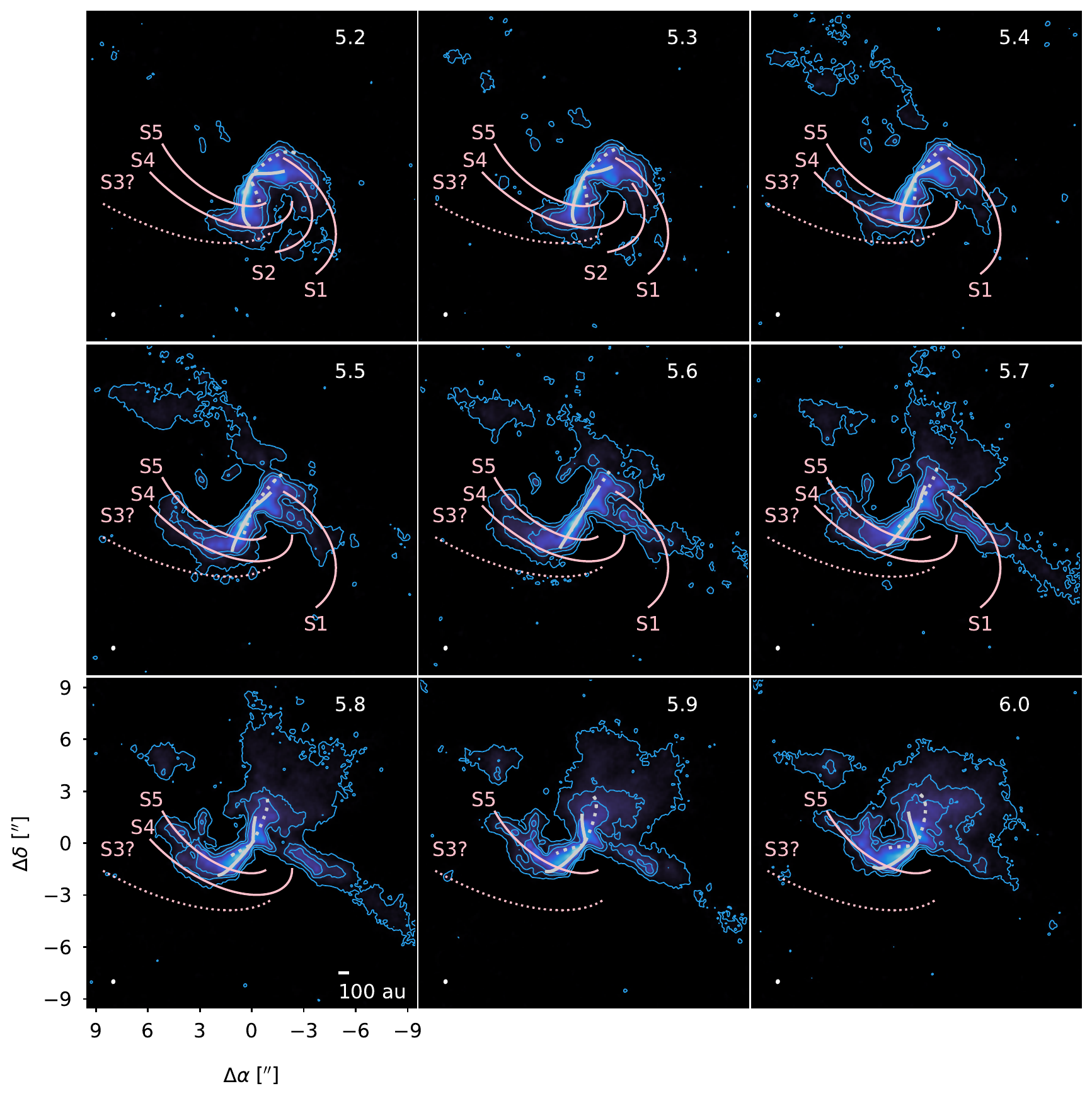}
\end{center}
\caption{Continued.}
\end{figure*}

Four spiral arms (S1, S2, S4, and S5) are identified based on visual inspection of GM Aur's $^{12}$CO $J=2-1$ channel maps, along with an additional tentative arm (S3). Figure \ref{fig:spiralchanmapspt1} shows the channel maps with logarithmic spiral curves overlaid (i.e., of the form $R(\varphi) = R_0 \exp{\left(b\varphi\right)}$, where $R$ is the distance of the spiral from GM Aur, $\varphi$ is the azimuthal angle in radians, and $R_0$ and $b$ are free parameters).  The coordinate system is defined to be in the plane of a geometrically thin disk with a position angle of $57\fdg17$ east of north and inclination of $53\fdg21$ \citep{2020ApJ...891...48H}. The northeast side of the disk major axis corresponds to $\varphi=\frac{\pi}{2}$, and $\varphi$ increases clockwise (see Figure 10 from \citet{2020ApJ...891...48H} for a schematic). Table \ref{tab:spirals} lists the parameter values of the overlaid curves, which were chosen to produce a reasonable visual match to the channel maps. A parametrization with constant pitch angle (i.e., a logarithmic spiral) is adopted for simplicity given the relatively low signal-to-noise ratio of the spiral arm emission (typically ranging from $5-10\sigma$ in a given location) and the small angular range over which the arms are detected (usually less than half a winding). 

\begin{deluxetable}{lllll}
\scriptsize
\tablecaption{Parameters of proposed spirals \label{tab:spirals}}
\tablehead{
\colhead{ID}&\colhead{$R_0$} &\colhead{ $b$} & \colhead{Pitch angle\tablenotemark{a}} & \colhead{$\varphi$ range\tablenotemark{b}} \\
&\colhead{(au)} & &\colhead{(Degrees)} & \colhead{(Radians)}}
\colnumbers
\startdata
S1& 1860  & 0.45 & 24 &($-0.9\pi$, $-0.3\pi$)\\
S2 &1235  & 0.4& 22& ($-0.7\pi$, $-0.2\pi$)\\
S3? & 1200 & 0.8 & 39 &($-0.2\pi$, $0.18\pi$)\\
S4 &825 &0.4&22& ($-0.5\pi$, $0.25\pi$)\\
S5&500&0.5 &27 & ($-0.25\pi$, $0.38\pi$)
\enddata
\tablenotetext{a}{The pitch angle of a logarithmic spiral is $\arctan b$.} 
\tablenotetext{b}{Approximate azimuthal range over which the spiral arm is visible in the $^{12}$CO channel maps.}
\end{deluxetable}

\begin{figure*}
\begin{center}
\includegraphics{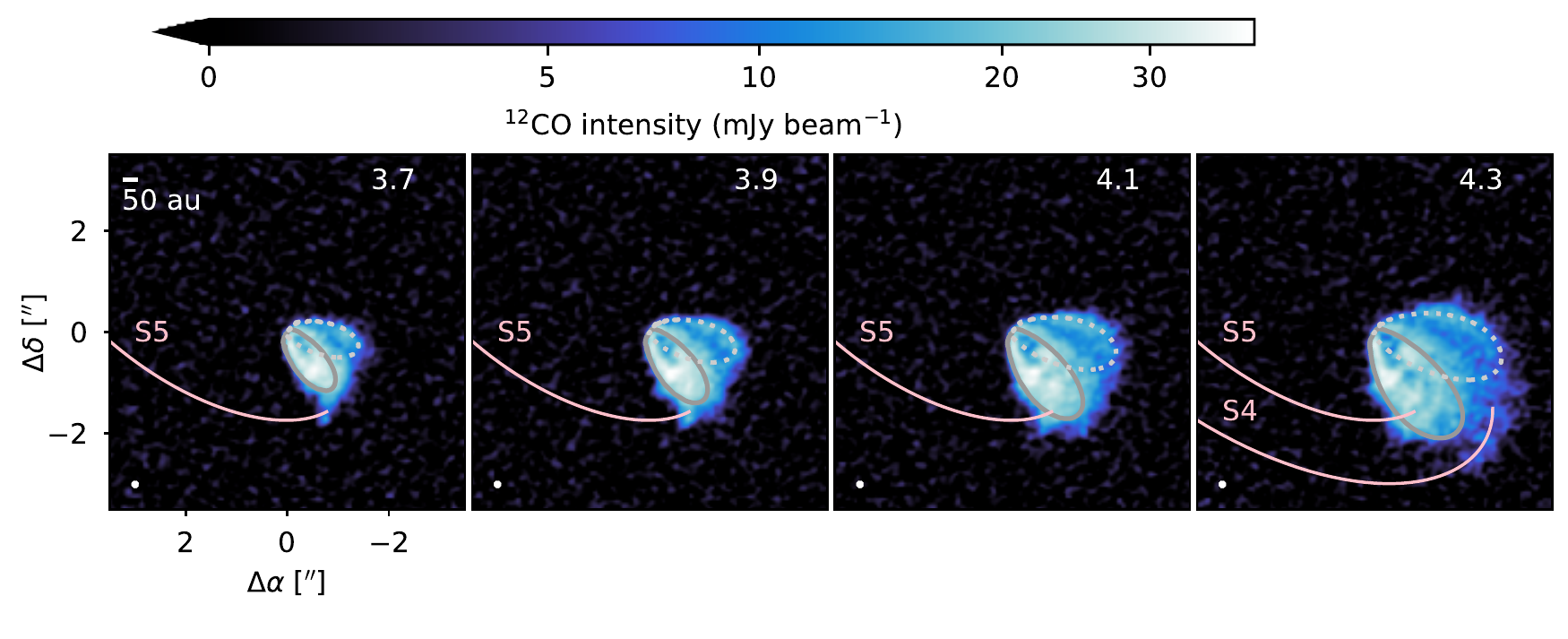}
\end{center}
\caption{Channel maps from the fiducial MAPS $^{12}$CO image, with curves for S4 and S5 overlaid to highlight the emission from these spirals at smaller disk radii. An arcsinh color stretch is used to help make the faint spiral features more visible. The spiral emission manifests as protrusions from the Keplerian disk. The isovelocity contours of the Keplerian disk are drawn in gray, with the solid curves tracing the front of the disk and the dotted curves tracing the back side. The upper right corner of each panel is labeled with the LSRK velocity (km s$^{-1}$). The synthesized beam ($0.15''\times0.15''$) is shown in the lower left corner of each panel. North is up and east is to the left. \label{fig:spiralchanmapshires}}
\end{figure*}

In general, a feature is considered to be detected if its emission exceeds the $4\sigma$ level in more than one channel. Nevertheless, because of the patchiness of some of the emission and blending of signal between the arms and with the Keplerian disk in some channels, reasonable alternative spiral identifications may be possible. In particular, S3 is the least certain because of the large discontinuities between detected emission features. While the tapering performed for the MAPS XIX version of the $^{12}$CO image generally makes the faint emission from the spiral arms more readily apparent in the channel maps, the lower resolution obscures some of the emission from S4 and S5 at smaller disk radii. The interior regions of these arms are more readily discernible in the higher resolution fiducial $^{12}$CO channel maps (Figure \ref{fig:spiralchanmapshires}).

\begin{figure}
\begin{center}
\includegraphics[scale=0.95]{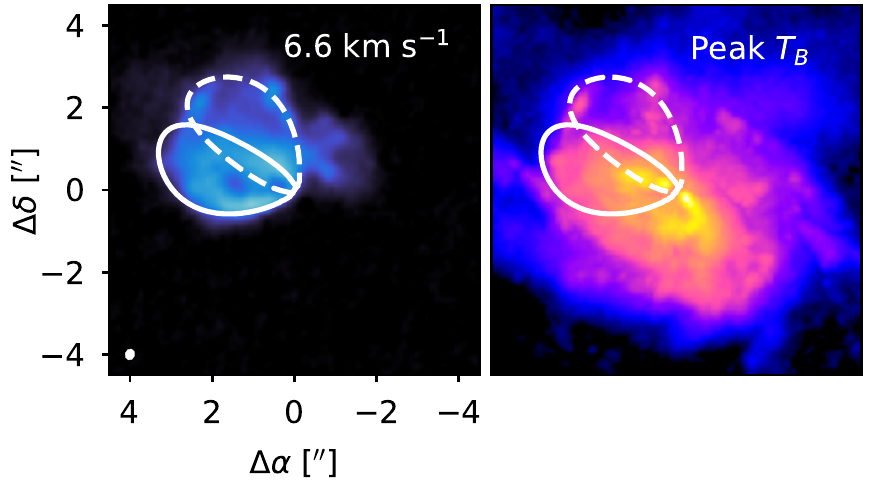}
\end{center}
\caption{A comparison of a $^{12}$CO channel map with the peak brightness temperature map suggests that the apparent arm-like structures to the northeast of the disk are due to emission from the disk's back side. Left: $^{12}$CO emission at 6.6 km s$^{-1}$, with the front and back sides of the Keplerian disk emission marked with solid and dashed contours, respectively. Right: $^{12}$CO peak brightness temperature map with the isovelocity contours corresponding to $v_{\text{lsr}}=6.6$ km s$^{-1}$ overlaid. The back side of the disk overlaps with the apparent arm-like structures.  \label{fig:back side}}
\end{figure}

In addition to the spiral arms identified in Table \ref{tab:spirals}, the $^{12}$CO peak brightness temperature map exhibits features resembling short arms on the northeast side of the disk. A comparison between the peak brightness temperature map and the $^{12}$CO channel maps, though, suggests that projection effects from the back side of the disk may be responsible for these arm-like emission features (Figure \ref{fig:back side}). 

Assuming that the spiral arms are in the plane of the disk, they are detected in $^{12}$CO at radii from $\sim340$ to $\sim$1200 au from the star, or $\sim$1900 au if the tentative S3 arm is included. Nevertheless, it is plausible that the spiral arms do not lie in the same plane as the disk or with one another. The analysis of the $^{12}$CO emitting height of the Keplerian portion of the disk in \citet{2021arXiv210906217L} indicates that the $J=2-1$ emission generally originates from well above the midplane ($z/r\sim0.4$), although the emitting height trends downward at larger radii and becomes highly uncertain at the projected distances of the spiral arms. Furthermore, the spiral arms deviate from Keplerian motion, as shown in Figure \ref{fig:spiralchanmapspt1}, where isovelocity contours for the Keplerian disk are drawn in gray for visual guidance. Details of how the isovelocity contours are computed, with adjustments for the pressure gradient, are provided in Appendix \ref{sec:isovelocities}. The non-Keplerian motions suggest that the arms could be oriented out of the plane of the disk. As demonstrated in \citet{2016ApJ...826...75D} and \citet{2020ApJ...900..135U}, projection effects can make even symmetric spiral arms appear to have dramatically different pitch angles and extents. Thus, while spiral arm parameters are provided for guidance, they should be interpreted with caution, e.g., variations in the listed pitch angles for different arms could in large part reflect uncertainties associated with the viewing angle.

\begin{figure*}
\begin{center}
\includegraphics[]{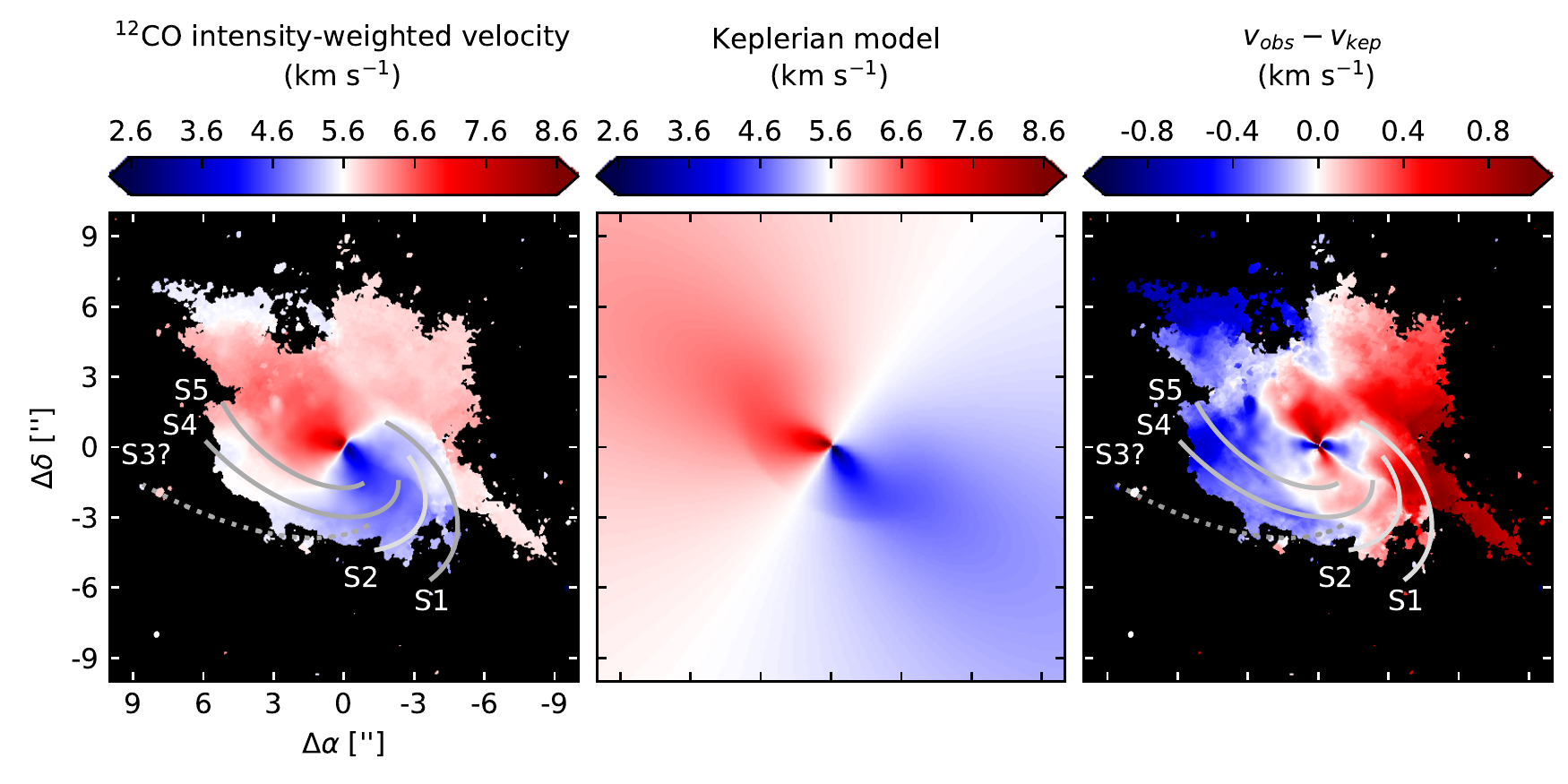}
\end{center}
\caption{Left: GM Aur's $^{12}$CO intensity-weighted velocity map, with the spiral curves from Table \ref{tab:spirals} overlaid in gray. Middle: A map of the expected Keplerian line-of-sight velocities corresponding to the front side of the GM Aur disk at the $^{12}$CO emitting height. Right: A map of the Keplerian velocity model subtracted from the observed intensity-weighted velocity map. \label{fig:velresiduals}}
\end{figure*}

The spiral arms' deviations from Keplerian motion are further highlighted in Figure \ref{fig:velresiduals} (this and all subsequent figures are based on the MAPS XIX image versions). In the intensity-weighted velocity map, the portion of S4 and S5 on the southeast side of the disk stand out as being blueshifted relative to the disk's Keplerian motion. To examine velocity deviations in more detail, a velocity residual map was created by computing a model Keplerian velocity field for the front side of the disk using the stellar properties, disk orientation, and $^{12}$CO emitting height described in Appendix \ref{sec:isovelocities}, then subtracting it from the intensity-weighted velocity map. Unlike the isovelocity contour calculations used for the channel maps, the pressure gradient is not included in the velocity residual calculations because Equation \ref{eq:vphi} yields imaginary solutions for $v_\varphi$ at large distances from the star (i.e., in the vicinity of the diffuse structures north of the disk), and would result in a discontinuous residual map. Overall, the residual map is quite complex not only because of the spiral arms, but also because of the southwest tail and diffuse northern emission. The non-Keplerian motion of the spiral arms can be traced to smaller radii in the velocity residual map compared to the intensity-weighted velocity map. In general, S4 and S5 are blueshifted relative to the expected Keplerian velocities, while S1 and S2 are redshifted. 

\begin{figure}
\begin{center}
\includegraphics{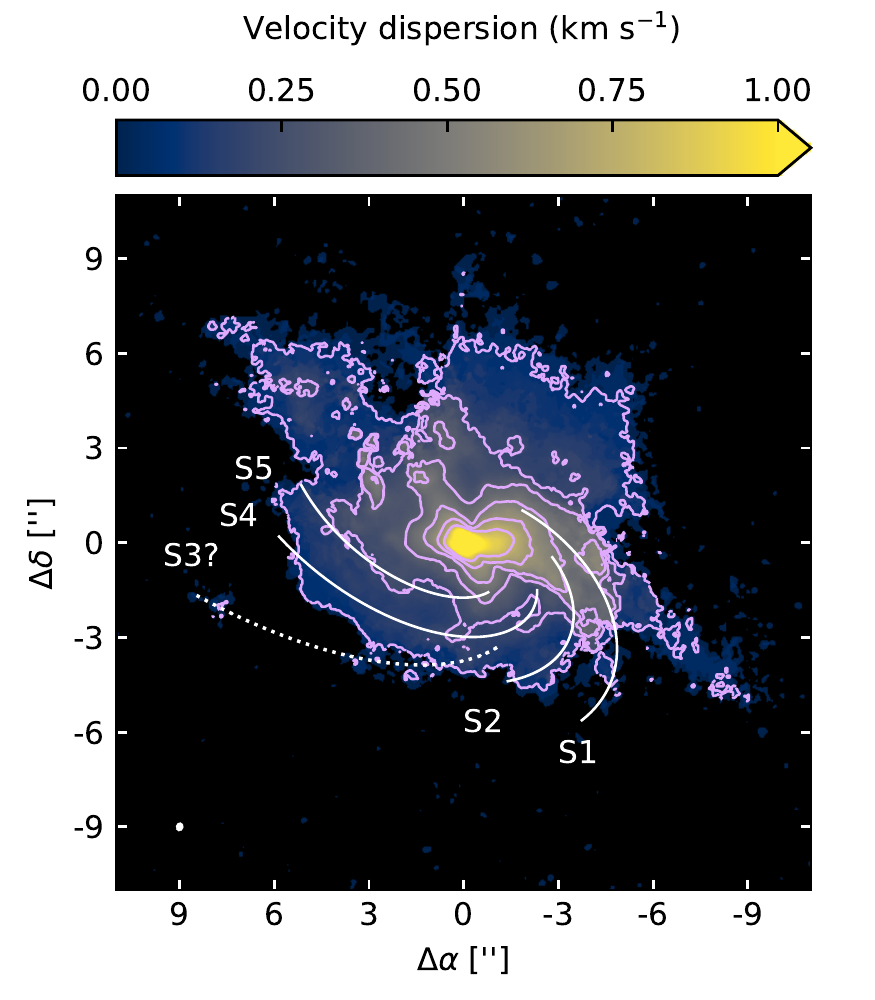}
\end{center}
\caption{Velocity dispersion map of $^{12}$CO toward GM Aur, with the spiral curves from Table \ref{tab:spirals} overlaid in light gray. Purple contours are drawn at 0.07, 0.21, 0.35, 0.49, 0.63, and 0.77 km s$^{-1}$.  \label{fig:velocitydispersion}}
\end{figure}

Figure \ref{fig:velocitydispersion} shows the $^{12}$CO velocity dispersion map, which was created with the CASA \texttt{immoments} task. Pixels below the $4\sigma$ level were clipped in order to avoid skewing from signal-free channels. For the most part, the spiral arms do not stand out prominently in their velocity dispersion, which is often computed to search for regions of increased turbulence. At smaller radii, S4 coincides with protrusions in the contour levels, which may hint at enhanced velocity dispersion in its vicinity. The southwest side of the disk, which overlaps with S1 and S2, exhibits more clearly elevated velocity dispersion, which can be attributed to the blending of multiple velocity components, including the two arms, the Keplerian disk, and the southwest tail (discussed further in Section \ref{sec:tail}).

\begin{figure*}
\begin{center}
\includegraphics[]{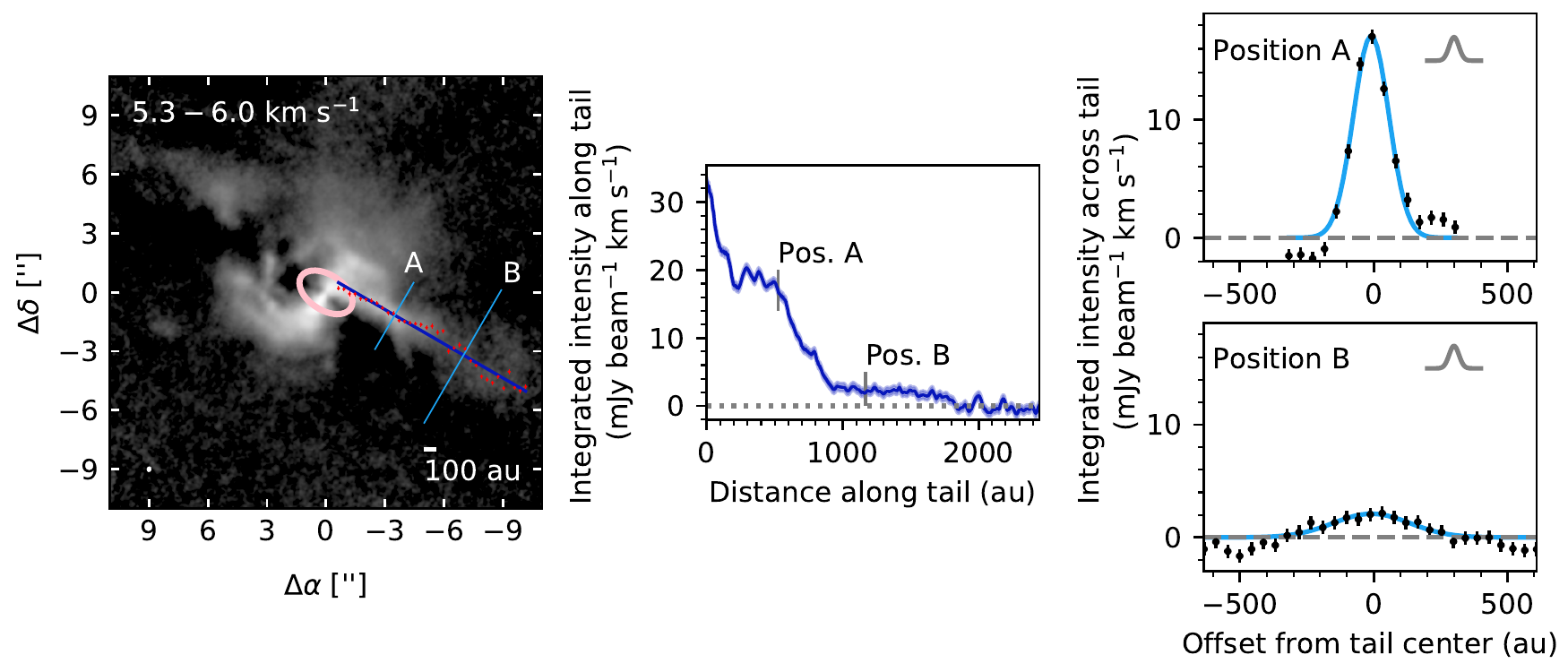}
\end{center}
\caption{Left: An integrated intensity map of $^{12}$CO emission from 5.3 to 6.0 km s$^{-1}$, highlighting emission from the southwest tail. The pink ellipse at the center of the image shows the orientation of the GM Aur disk. The red points show the positions of local maxima measured in vertical slices of the image, along with $1\sigma$ error bars. The dark blue line segment shows the best fit to the red points. The thin, light blue lines show the positions at which intensities are measured across the width of the tail. Middle: The integrated intensity of $^{12}$CO measured along the best-fit line to the tail, starting from the northeast end. The lavender shading shows the $1\sigma$ noise level. Right: The intensities measured along positions A and B, respectively. The measurements and $1\sigma$ error bars are plotted in black. The intensity along position A is measured within a narrower interval in order to exclude emission from other disk structures. The offset from tail center is defined with respect to the best-fit line segment drawn in the left-hand panel, with negative offsets southeast of the line and positive offsets northwest. The light blue curves show the best-fit Gaussians to the measured intensity profiles. The gray Gaussian curves in the upper right-hand side of each plot show the width of the synthesized beam.   \label{fig:tailplots}}
\end{figure*}

\subsubsection{The southwest tail}\label{sec:tail}
The southwest tail is visible in the $^{12}$CO channel maps from $\sim5.3-6.0$ km s$^{-1}$ and $6.3-6.8$ km s$^{-1}$. In between these velocity ranges, only broad emission is observed, implying either that the diffuse northern emission component is in front of the tail at these velocities or that they are blended due to similar brightnesses. To characterize the geometry of the tail, we produced an integrated intensity map from $\sim5.3-6.0$ km s$^{-1}$, without applying any intensity cutoffs (Figure \ref{fig:tailplots}). The higher velocity channels are excluded in order to avoid confusion with the diffuse northern emission, but this exclusion does not substantially affect characterization of the tail's geometry because the tail emission at higher velocities is compact. The rms, measured in a signal-free region of this map, is 0.6 mJy beam$^{-1}$ km s$^{-1}$. To measure the orientation of the tail, we took a series of vertical cuts across the image spaced $0\farcs28$ apart (i.e., comparable to the synthesized beam) in the east-west direction and identified local intensity maxima along these cuts. We consider the starting point of the tail to be $0\farcs64$ west of GM Aur, which is the closest point to the star at which the tail can be distinguished from the Keplerian disk emission via separate local maxima in a vertical slice across the image. Vertical cuts were taken until the local maximum fell below $3\sigma$, which occurs $10\farcs2$ west of GM Aur. The uncertainty in the position of each local maximum is taken to be the standard deviation of the Gaussian synthesized beam. Since the tail does not exhibit substantial curvature, we fit a line, $y = m_{\text{tail}} x+b_{\text{tail}}$, to the positions of all the local maxima using least-squares minimization. The variables $x$ and $y$ denote the offsets (in arcseconds) from the center of the GM Aur disk (which is aligned with the phase center) in the eastern and northern directions, respectively. The best-fit tail parameter values are $m_{\text{tail}}=0.580\pm0.007$ and $b_{\text{tail}}=0\farcs86\pm0\farcs04$. The lefthand panel of Figure \ref{fig:tailplots} shows the best-fit line segment plotted over the integrated intensity map of the tail. The best-fit line segment is oriented $59\fdg9\pm0.3$ east of north, which differs only by a few degrees from the GM Aur disk's  position angle of $57\fdg17\pm0.02$ \citep{2020ApJ...891...48H}. The end of the best-fit line segment closest to GM Aur is located $\sim0\farcs8$ northwest of the star. If the tail is co-planar with the disk, then the tail extends as far inward as $\sim200$ au from GM Aur.

The middle panel of Figure \ref{fig:tailplots} shows the $^{12}$CO integrated intensity measured along the best-fit line segment. The measured intensity is highest in the inner tail, although it cannot be fully disentangled from disk emission. The integrated intensity decreases steeply with distance up to $\sim200$ au along the tail. From $\sim200$ to $600$ au, the intensity appears to be relatively flat, perhaps with some small amplitude variations. The intensity then decreases steadily until $\sim900$ au from the start of the tail, at which point it abruptly plateaus. Faint tail emission continues to be visible up to a distance of $\sim1800$ au. 

In projection, the tail widens modestly with distance from GM Aur. The righthand panel of Figure \ref{fig:tailplots} shows the integrated intensity measured along two perpendicular cuts across the tail. Position A is located 526 au from the start of the tail, while position B is located 1169 au from the start. These positions were chosen to be well-separated from each other and to avoid blending with disk emission. Along each position, intensity measurements were taken at intervals of $0\farcs28$ (i.e., spaced one synthesized beam apart). The width of the tail at those two positions was estimated by fitting a Gaussian, $I(d) = I_0\exp\left(-0.5(d-d_0)^2/\sigma^2 \right)$, where $d$ is the perpendicular offset from the best-fit line along the length of the tail. The best fit parameters are  $I_0=17.1\pm0.5$ mJy beam$^{-1}$ km s$^{-1}$, $d_0 = -10\pm2$ au, and $\sigma=67\pm2$ au at position A and $I_0=2.1\pm0.3$ mJy beam$^{-1}$ km s$^{-1}$, $d_0 = -10\pm20$ au, and $\sigma=140\pm20$ au at position B. The systematic flux calibration uncertainty contributes another $\sim10\%$ uncertainty in $I_0$. Spatial filtering may lead to underestimates in the width of the tail, but in any case, it is spatially resolved.

\subsubsection{Diffuse northern emission}\label{sec:diffuse}
\begin{figure*}
\begin{center}
\includegraphics{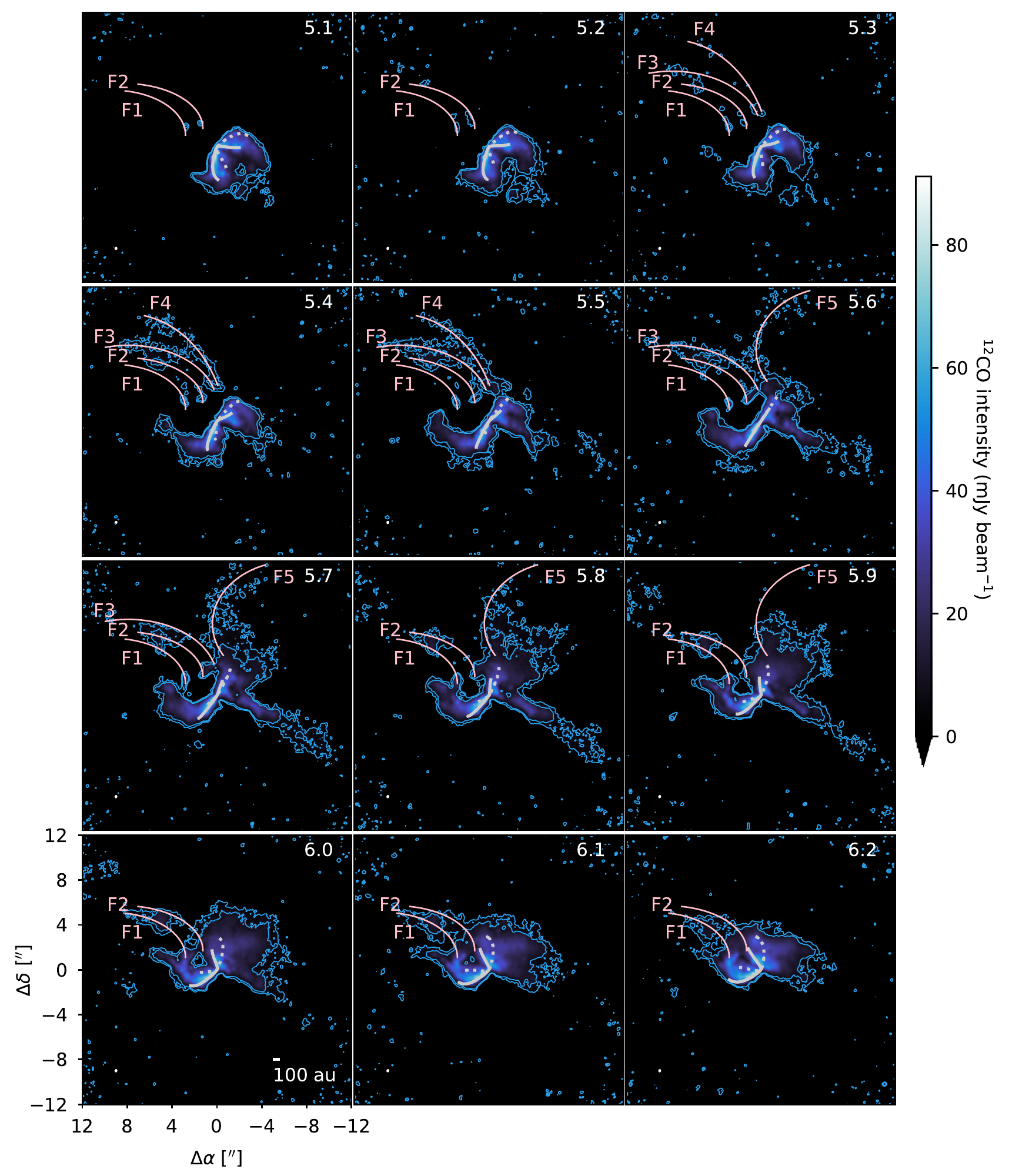}
\end{center}
\caption{$^{12}$CO channel maps toward GM Aur with pink curves overlaid to mark tentative identifications of filamentary components within the diffuse emission north of the disk. The light blue contours show the 4 and 8$\sigma$ emission levels, where $\sigma=1$ mJy beam$^{-1}$. The isovelocity contours of the Keplerian disk are drawn in light gray, with the solid curves tracing the front of the disk and the dotted curves tracing the back side. The upper right corner of each panel is labeled with the LSRK velocity (km s$^{-1}$). Synthesized beams are shown in the lower left corner of each panel. \label{fig:arcchanmaps}}
\end{figure*}

Diffuse emission north of GM Aur is present in the $^{12}$CO channel maps near the disk systemic velocity (5.61 km s$^{-1}$). Five filamentary components (F1, F2, F3, F4, and F5) are tentatively identified within this diffuse emission and highlighted with elliptical arcs in Figure \ref{fig:arcchanmaps}. The elliptical arcs are parametrized as $x = A\cos t + x_0$, $y = B\sin t + y_0$, where $x$ and $y$ are the offsets (in arcseconds) east and north of GM Aur, respectively, and $t$ is an angular variable in radians. This parametrization is selected because it reasonably approximates the morphology of the filamentary components, but other parametrizations may be similarly suitable. Table \ref{tab:filaments} lists the parameter values for each filamentary component, which are determined via visual inspection. As with the spiral structures, the patchiness of the diffuse northern emission creates some ambiguity as to which structure individual emission components should be assigned. Furthermore, a possible concern in the interpretation of the CO observations is that spatial filtering may be inducing the appearance of filamentary substructure within the diffuse northern emission. However, the identification of some of these filamentary components is supported by their alignment with structures observed in scattered light (see Section \ref{sec:hst}).

 \begin{deluxetable}{llllll}
 \scriptsize
 \tablecaption{Parameters of elliptical arcs used to make PV diagrams of tentative filament structures}\label{tab:filaments}
 \tablehead{\colhead{ID}&\colhead{$A$} &\colhead{ $B$} & \colhead{$x_0$} & \colhead{$y_0$}&\colhead{$t$ range\tablenotemark{a}}\\
 \colhead{}&\colhead{(arcsec)} &\colhead{(arcsec)} & \colhead{(arcsec)} & \colhead{(arcsec)}&\colhead{(radians)}}
 \startdata
 F1&6.5&4 &9.3 & 1.1 &($\frac{5\pi}{9}$, $\pi$) \\
 F2&7&4 &8.25 & 1.7 &($\frac{5\pi}{9}$, $\pi$)  \\
 F3&7.9&5.3 &7.9 & 1.5 &($\frac{5\pi}{12}$, $\frac{11\pi}{12}$)  \\
 F4&8.9&11 &8 & $-1.4$ &($\frac{5\pi}{9}$, $\frac{31\pi}{36}$) \\
 F5 &7.2&5.6 &$-6.8$ & 6.4 &($-\frac{\pi}{6}$, $\frac{7\pi}{18}$)
 \enddata
\tablenotetext{a}{$t=0$ points east and $t=\pi/2$ points north.}
 \end{deluxetable}

\begin{figure}
\begin{center}
\includegraphics{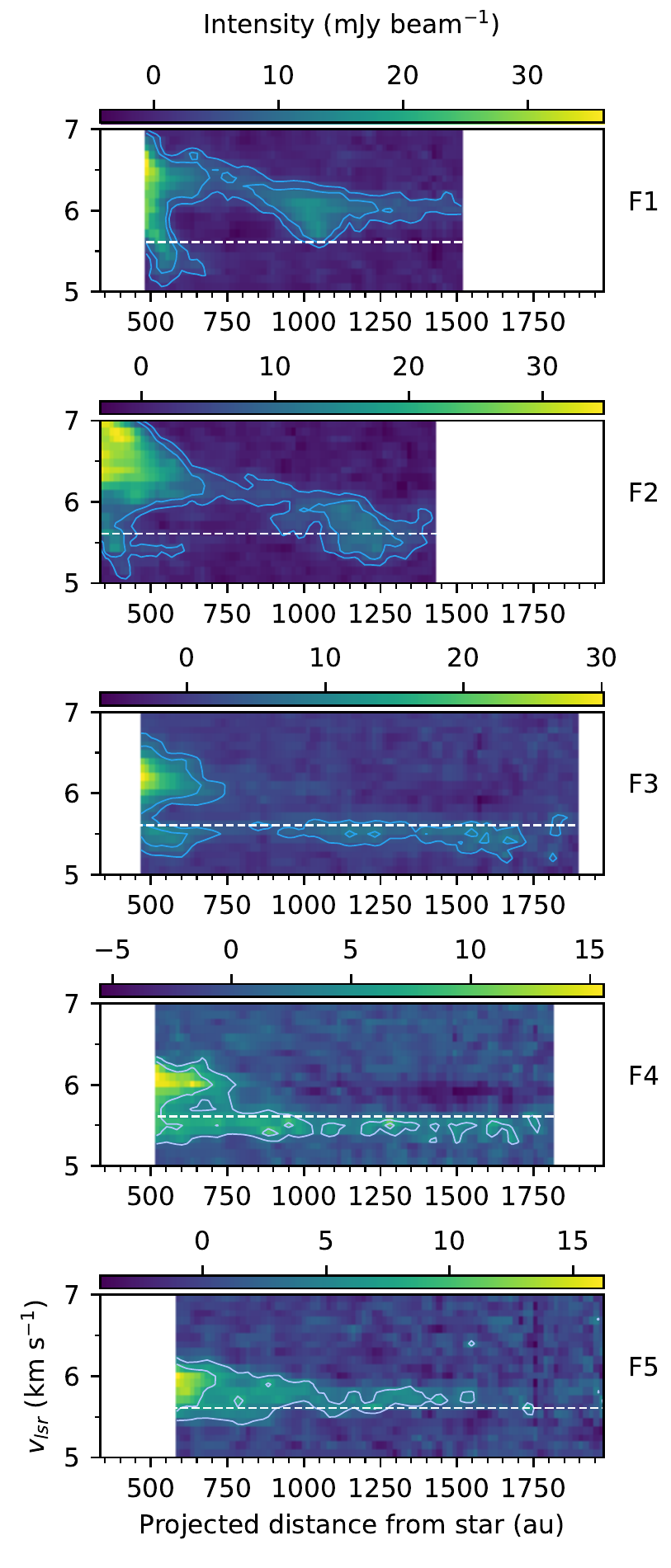}
\end{center}
\caption{Position-velocity diagrams of the tentatively identified filaments in the diffuse northern emission, extracted along the elliptical arcs drawn in Figure \ref{fig:arcchanmaps}. Contours are drawn at the 4 and 8 mJy beam$^{-1}$ levels (corresponding to 4 and 8$\sigma$ in the channel maps). The white dashed line marks GM Aur's systemic velocity.  \label{fig:filamentvelocities}}
\end{figure}

\begin{figure}
\begin{center}
\includegraphics{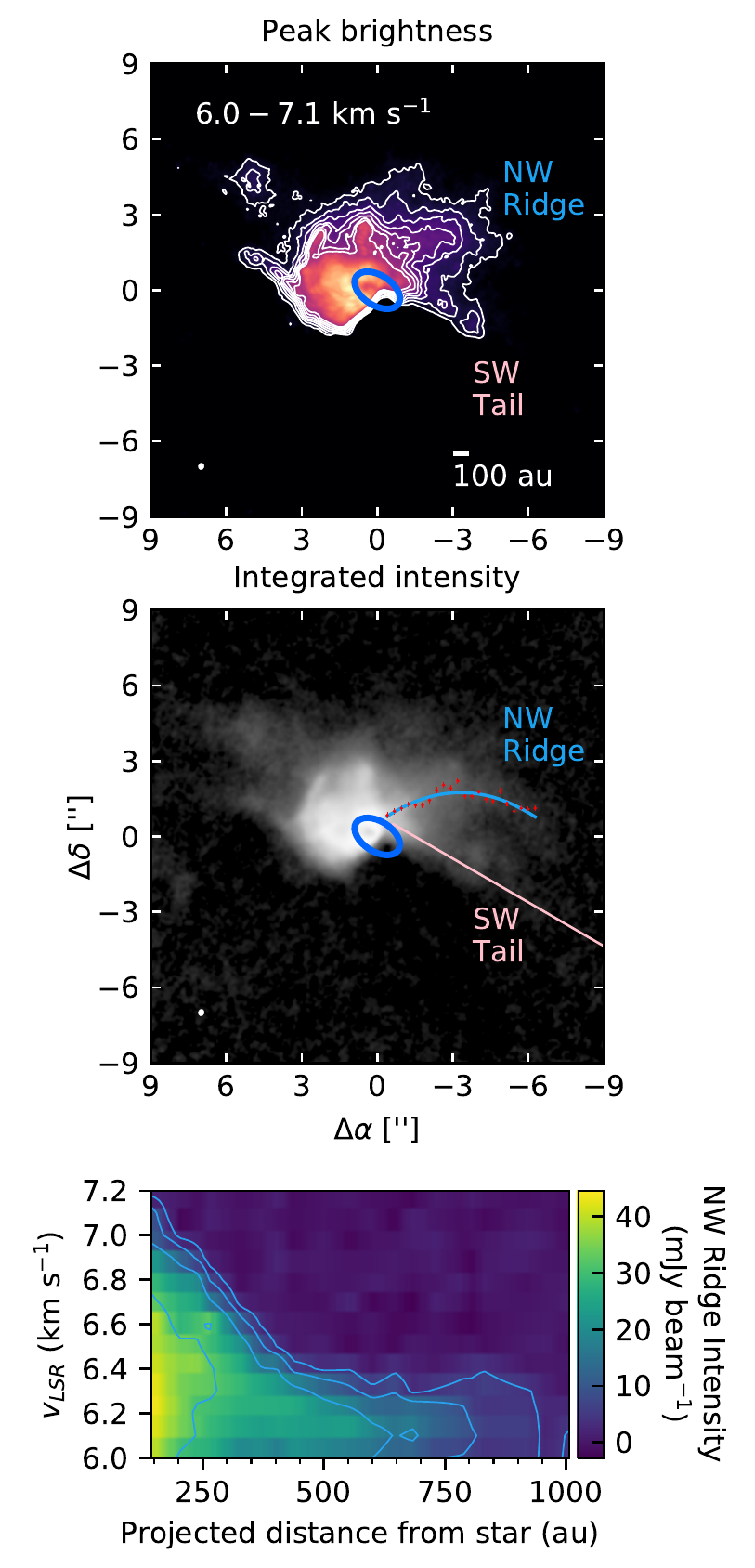}
\end{center}
\caption{Top: A $^{12}$CO peak brightness map from channels between 6.0 and 7.1 km s$^{-1}$, showing a broad ``ridge'' of emission northwest of the disk. The collimated component to the south is part of the southwest tail. Contours are drawn in increments of 5 mJy beam$^{-1}$, starting from 10 and ending at 40 mJy beam$^{-1}$. The blue ellipse shows the position angle and inclination of the GM Aur disk. Middle: The integrated intensity map from 6.0 to 7.1 km s$^{-1}$, with the measured positions of the local maxima along the northwest ridge and the $1\sigma$ error bars plotted in red. The best fit quadratic curve to the shape of the northwest ridge is drawn in light blue. The southwest tail is marked in pink. Bottom: Position-velocity diagram extracted along the best-fit curve to the northwest ridge. Contours are drawn at [4, 8, 16, 32] mJy beam$^{-1}$. \label{fig:westernarc}}
\end{figure}

Figure \ref{fig:filamentvelocities} shows position-velocity diagrams extracted along the arcs drawn in Figure \ref{fig:arcchanmaps}. In general, F1 and F2 exhibit decreasing offsets from the systemic velocity with increasing projected distance from GM Aur. In other words, at larger projected distances, the gas in F1 and F2 is moving more slowly along the line of sight relative to GM Aur. Meanwhile, F3, F4, and F5 do not exhibit much variation in velocity with projected distance from the star. F3 and F4 are slightly blueshifted with respect to $v_{\text{sys}}$, while F5 is slightly redshifted. The Keplerian disk is responsible for the bright second emission component visible at a projected distance of $\sim500$ au and 6 km s$^{-1}$ in the diagrams for F3 and F4. At projected distances less than $\sim700$ au, emission from F1 and F2 is spread over a wide velocity range. The Keplerian disk  contributes to some of the bright emission visible at smaller projected distances and (relatively) large velocities in the diagrams for F1 and F2, but as can be seen in Figure \ref{fig:arcchanmaps}, much of the emission visible at the smaller projected distances originates from clumps at the inner tips of F1 and F2. (We refer to these as clumps based on their appearance in the channel maps, but note that the clumpiness may be due to the relatively low signal-to-noise ratio or imaging artifacts rather than the intrinsic physical properties).

The diffuse emission northwest of the disk is mostly redshifted relative to the systemic velocity, and does not exhibit as much substructure as the emission northeast of the disk. The brightest emission within the northwest component traces a broad arc (which is subsequently referred to as the ``northwest ridge''), as shown in the peak brightness map created from channels between 6.0 and 7.1 km s$^{-1}$ and with a $5\sigma$ clip (Figure \ref{fig:westernarc}). The shape of the ridge was measured in a manner similar to the southwest tail, as described in Section \ref{sec:tail}. We created an integrated intensity map from 6.0 to 7.1 km s$^{-1}$ (with no pixels clipped) and took a series of vertical cuts across the image spaced $0\farcs28$ apart (i.e., roughly one synthesized beam) in the east-west direction. Local intensity maxima were identified along these cuts between $0\farcs40$ to $6\farcs28$ west of the star, and their positions relative to GM Aur were fit with a quadratic, $y = a_{\text{NW Ridge}}x^2+b_{\text{NW Ridge}}x+c_{\text{NW Ridge}}$, where the positive $x$ and $y$ directions are east and north of the star, respectively. We use this parametrization because a second-order polynomial is the simplest curve that approximates the shape of the northwest ridge. Using least-squares minimization, we find that the best fit is $a_{\text{NW Ridge}} = -0.109\pm0.009$ arcsec$^{-1}$, $b_{\text{NW Ridge}} =-0.73\pm0.06$, and $c_{\text{NW Ridge}} = 0.53\pm0.09$ arcsec. Figure \ref{fig:westernarc} shows the best fit curve overlaid on the integrated intensity map. 

A position-velocity diagram extracted along the best-fit curve to the northwest ridge is shown at the bottom of Figure \ref{fig:westernarc}. At smaller velocities and projected distances within a few hundred au from GM Aur, some of the emission along the curve comes from the Keplerian disk (see Figure \ref{fig:northwestchanmaps}). At velocities higher than 6.3 km s$^{-1}$, though, the diagram traces only emission associated with the ridge. As the velocity increases, the emission extent of the ridge in individual channels becomes more compact. Although the diffuse northern emission had previously been ascribed to cloud contamination by \citet{2009ApJ...698..131H}, the tendency for the diffuse structures to have higher line-of-sight velocities at smaller projected distances from GM Aur suggests that gas is moving faster closer to the star, which is consistent with the gas motion being influenced by GM Aur's gravity. Thus, while the extended emission may technically originate from cloud material, the kinematics imply close interaction with the GM Aur system. 

\begin{figure*}
\begin{center}
\includegraphics{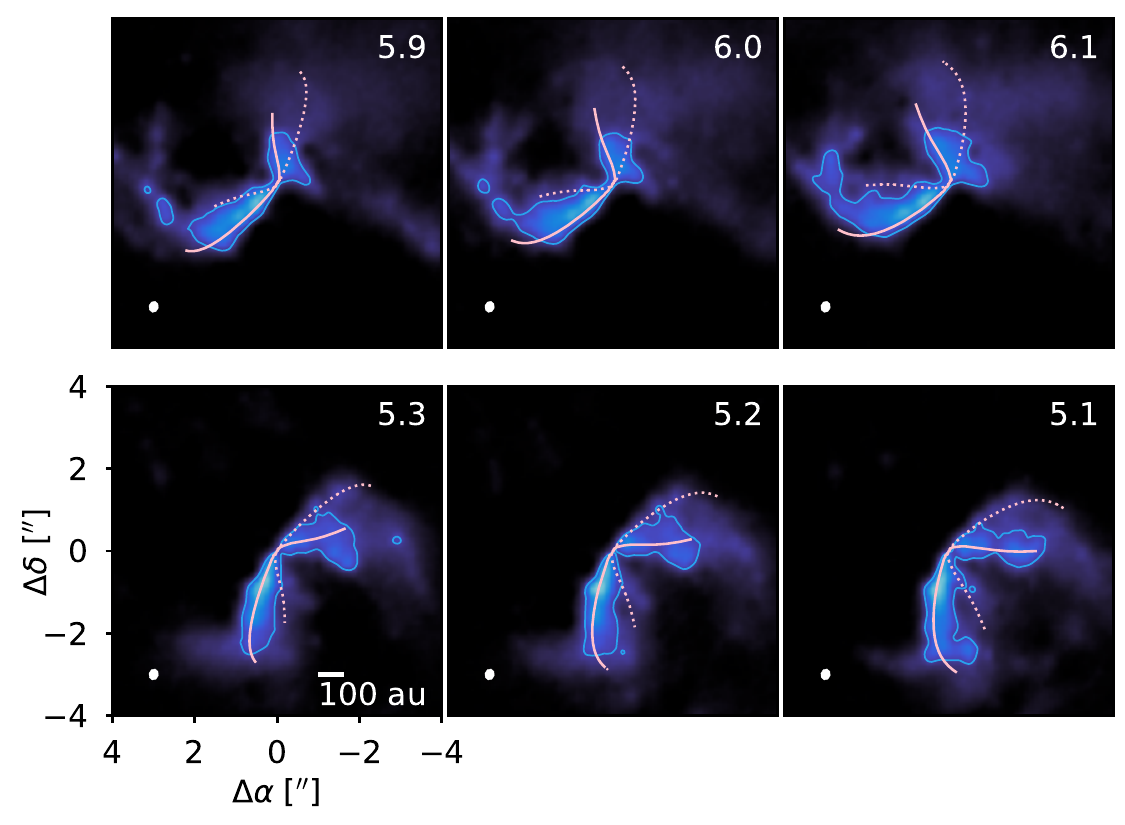}
\end{center}
\caption{Top row: In several channels redshifted from $v_{\text{sys}}$, the northern portion of the Keplerian disk is partially obscured by diffuse emission structures. The blue contour shows the $40\sigma$ emission level. The Keplerian isovelocity contours are drawn in pink, with the solid curves denoting the front of the disk and the dotted curves denoting the back side. The upper right corner of each panel is labeled with the LSRK velocity in km s$^{-1}$. Bottom row: The corresponding blueshifted channels are shown, demonstrating that the northern portion of the Keplerian disk extends further than what is visible in the redshifted channels. \label{fig:foregroundchanmaps}}
\end{figure*}

At several of the channels redshifted relative to $v_{\text{sys}}$, the full extent of the northern portion of the Keplerian disk is not visible (Figure \ref{fig:foregroundchanmaps}). To demonstrate the difference from the expected appearance of the Keplerian disk, Figure \ref{fig:foregroundchanmaps} also shows the corresponding blueshifted channels, which would otherwise be roughly mirror images across the disk minor axis (with some small deviations due to spiral structures in the outer disk). We interpret the truncated appearance of the northern side of the Keplerian disk in redshifted channels to be due to foreground obscuration by cooler, optically thick CO gas in the diffuse northwest emission structures. The relative positions and velocities of the diffuse structures and the disk thus imply that the large-scale northwest gas is moving toward the disk; i.e., it is infalling.

\begin{figure*}
\begin{center}
\includegraphics[]{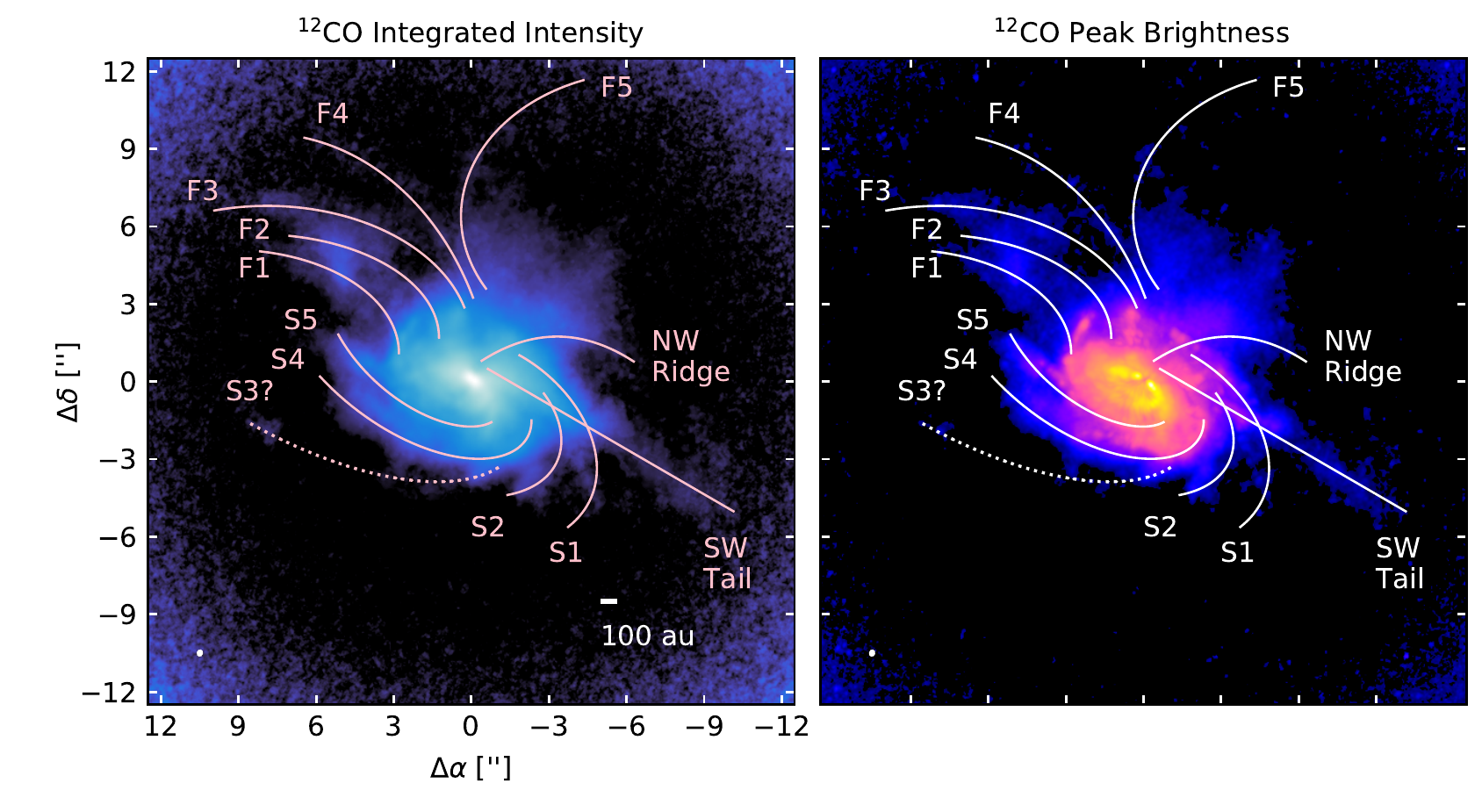}
\end{center}
\caption{Maps of the $^{12}$CO integrated intensity (left) and peak brightness (right), annotated with the locations of the spiral arms, diffuse northern emission components, and the southwest tail. The ``vignetting'' effect at the edges of the integrated intensity map arises from the decreasing sensitivity away from the phase center due to the primary beam. \label{fig:annotatedmommaps}}
\end{figure*}

\subsubsection{Summary of CO features}

Figure \ref{fig:annotatedmommaps} illustrates the positions of all identified $^{12}$CO features in relation to one another. While portions of these features are readily visible in the integrated intensity and peak brightness maps, others required inspection of channel maps and velocity maps in order to ascertain their positions. In projection, these features exhibit a wide range of orientations, morphologies, and extents. Although features are labelled differently depending on their appearance (e.g., the northwest ridge is broader than the northeast filamentary structures and the southwest tail doesn't exhibit the curvature of the other structures), viewing angle may be responsible for some of this morphological variation. Thus, different labels do not necessarily imply different physical origins. In addition, the relative positions of some features identified as separate structures could instead be part of a single larger structure, with the emission from the Keplerian disk or lack of sensitivity creating the appearance of separate features. In particular, the orientations of the southwest tail, F1, F2, and F3 are suggestive because they all lie roughly along the disk major axis. Finally, several of the structures appear to extend close to the FWHM of ALMA's primary beam ($\sim27''$ at 230 GHz). Given the dropoff in sensitivity with distance from the phase center, it is likely that our data do not recover the full extent of these structures. In the future, mosaicked observations and better $uv$ coverage will be key for fully characterizing the complexity of GM Aur's immediate environment. 

\subsection{Comparison to HST scattered light observations }\label{sec:hst}

\begin{figure*}
\begin{center}
\includegraphics[]{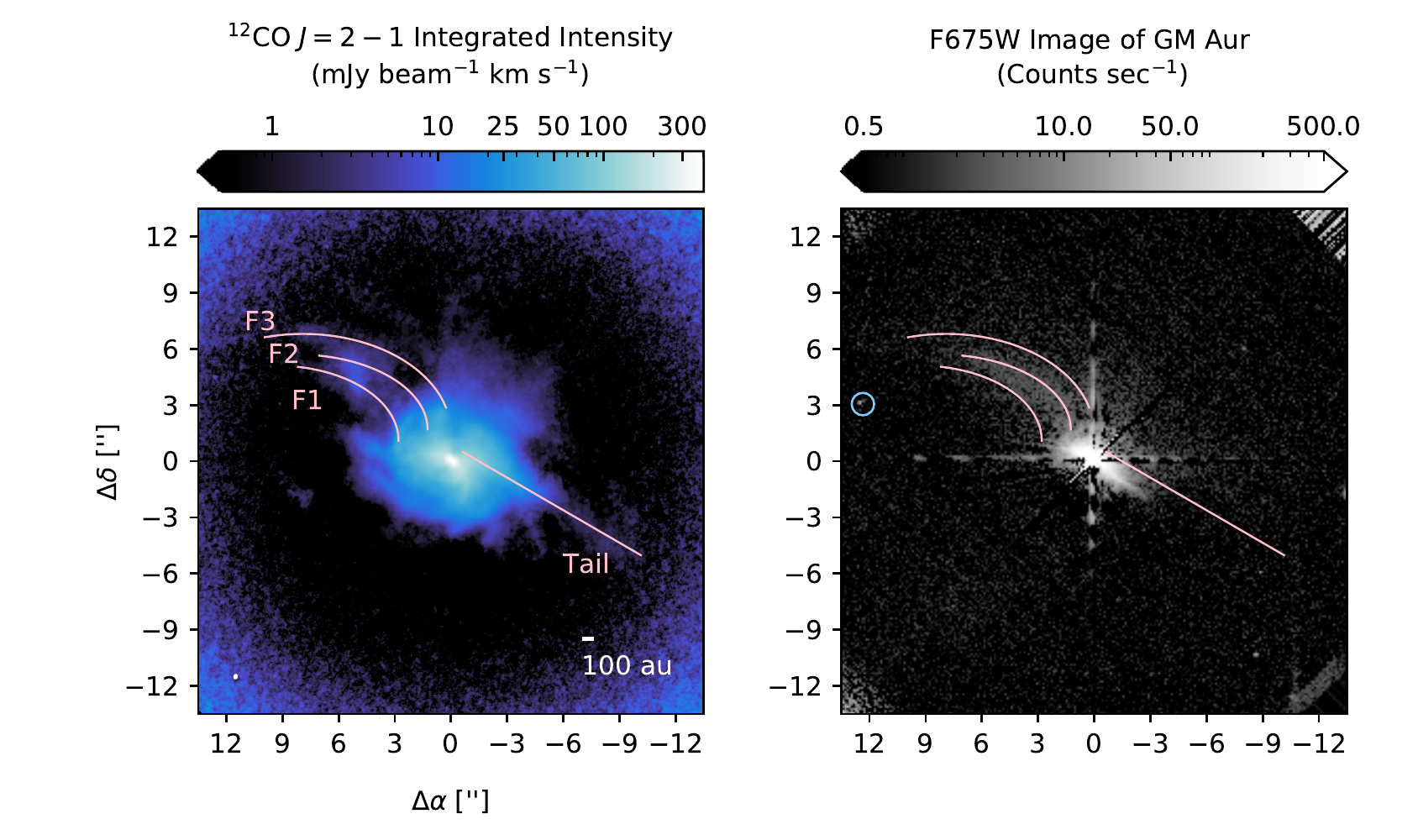}
\end{center}
\caption{A comparison of GM Aur's $^{12}$CO $J=2-1$ integrated intensity map to the WFPC2 F675W optical image. The pink curves mark the $^{12}$CO substructures that occur in close proximity to scattered light features. The blue circle in the WFPC2 image marks the location of USNO-B1.0 1203-0070338. The linear artifacts intersecting with the disk are diffraction spike residuals. North is up and east is to the left. Both images are shown with a logarithmic color stretch to highlight the faint extended features. \label{fig:wfpc2comparison}}
\end{figure*}

While none of the extended $^{12}$CO structures have counterparts in millimeter continuum emission, which nominally traces millimeter-sized dust grains in the disk midplane, some of the structures have counterparts in optical and near-infrared scattered light images, which trace sub-micron-sized dust grains that are better coupled to the gas. Figure \ref{fig:wfpc2comparison} compares the $^{12}$CO integrated intensity map to the WFPC2 F675W image of GM Aur. F1 and F3 follow the southern and northern edges, respectively, of the elongated scattered light feature northeast of the disk, while F2 falls roughly along the middle of the feature. This scattered light feature was previously detected in R-band with the University of Hawaii 2.2 m telescope \citep{Kalasthesis} and at 1.1 and 1.6 $\mu$m with HST NICMOS \citep{2003AJ....125.1467S}. \citet{2003AJ....125.1467S} named this feature the ``blue ribbon.'' Substructure within the northeast dust ribbon (e.g., distinguishing F2 from F1 and F3) is not clearly visible in any of these scattered light images.

A point source, marked with a blue circle, is visible slightly to the southeast of the dust ribbon in the F675W image shown in Figure \ref{fig:wfpc2comparison}. \citet{Kalasthesis} suggested that a close encounter between the point source and GM Aur was responsible for the dust ribbon, given their apparent proximity. A search of the Vizier database indicates that the point source coincides with USNO-B1.0 1203-0070338, located at J2000 04:55:11.894+30:22:03.032 \citep{2003AJ....125..984M}. This source does not appear in the Gaia database \citep{2016AA...595A...2G, 2018AA...616A...1G}, so no distance estimate appears to be available. However, the proper motion reported in the USNO-B catalog ($\mu_\alpha=-152\pm18$ mas yr$^{-1}$, $\mu_\delta=-124\pm129$ mas yr$^{-1}$) is inconsistent with the median proper motion measured by \citet{2018AJ....156..271L} for the L1517 cloud ($\mu_\alpha=4.7\pm0.8$ mas yr$^{-1}$, $\mu_\delta=-24.5\pm1.7$ mas yr$^{-1}$), in which GM Aur is located. Thus, we consider it unlikely that USNO-B1.0 1203-0070338 is physically in close proximity to GM Aur.

\begin{figure*}
\begin{center}
\includegraphics[]{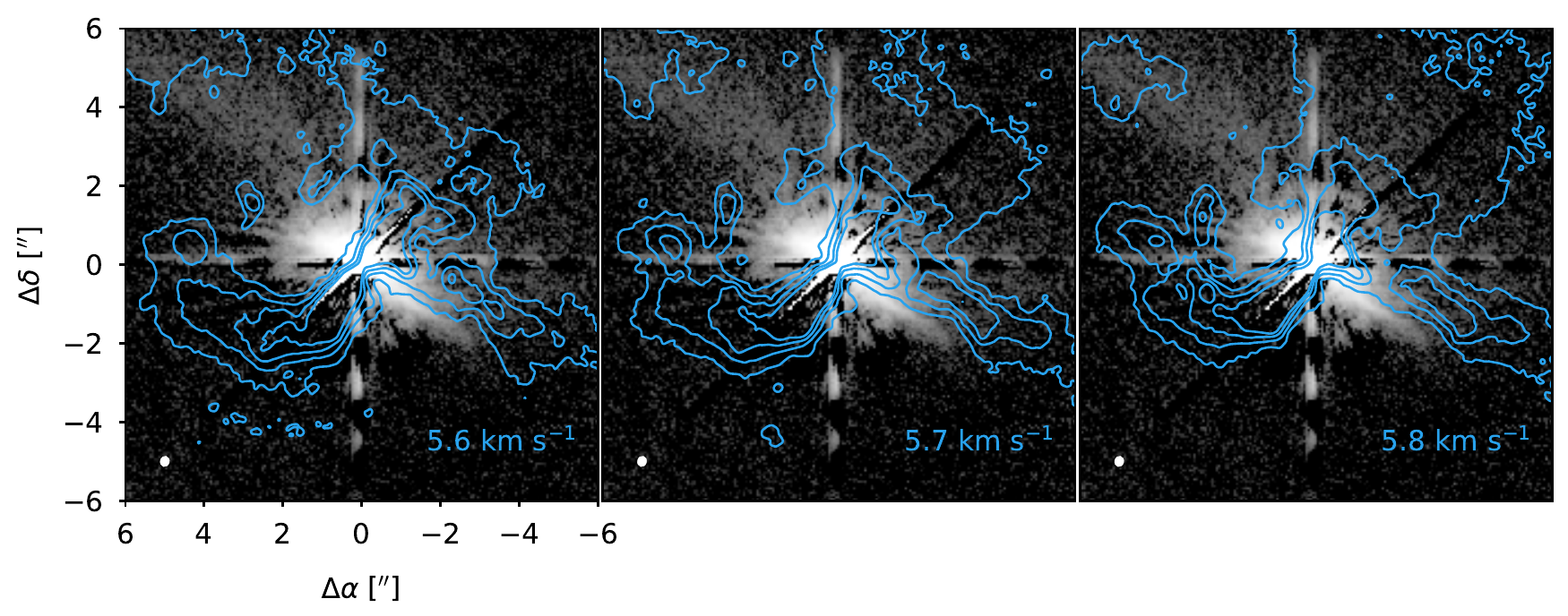}
\end{center}
\caption{The WFPC2 F675W image of GM Aur (grayscale) overlaid with $^{12}$CO channel map contours (5, 15, 25, and 35$\sigma$). The tail-like structure to the southwest of GM Aur in scattered light flanks the southeastern edge of the tail traced by $^{12}$CO emission. The $^{12}$CO LSRK velocity is labeled in the lower right corner of each panel. North is up and east is to the left. \label{fig:wfpc2overlay}}
\end{figure*}

Faint nebulosity is visible northwest of the disk in the WFPC2 image, overlapping with some of the diffuse $^{12}$CO emission. A bright tail-like structure appears to the southwest of the disk. This feature is also visible in a WFPC2 F555W image published in \citet{2016ApJ...829...65H}, although it was not specifically described as a tail. The scattered light tail is offset from where the tail is brightest in $^{12}$CO emission. Instead, as shown in a comparison between the WFPC2 F675W image and several of the $^{12}$CO channels, the scattered light tail flanks the southeastern edge of the CO emission tail (Figure \ref{fig:wfpc2overlay}). This visual offset may be the consequence of an offset in the $\tau=1$ surfaces, but radiative transfer modeling would be needed to confirm. The $^{12}$CO tail contours appear to bend slightly at the outer tip of the scattered light tail, suggesting some change in the physical properties. Overall, the correspondence between scattered light and $^{12}$CO structures indicates that the general morphologies of the complex gas features are not significantly distorted by spatial filtering.

\begin{figure*}
\begin{center}
\includegraphics[]{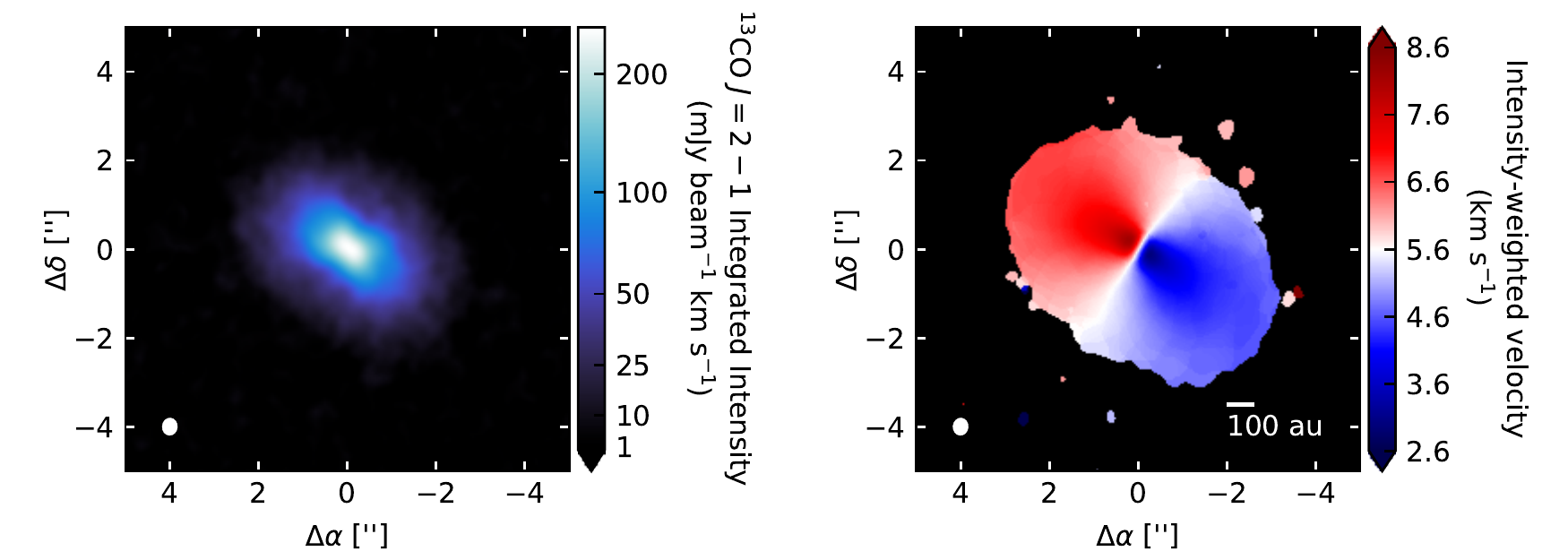}
\end{center}
\caption{The integrated intensity (left) and intensity-weighted velocity (right) maps for $^{13}$CO $J=2-1$ toward GM Aur. A logarithmic color stretch is used for the integrated intensity map to make the emission in the outer disk more visible. The synthesized beam size, $0\farcs39\times0\farcs35$ (-4\fdg4), is represented with a white ellipse in the lower left corner of each panel. \label{fig:13COmommaps}}
\end{figure*}
\subsection{Tentative $^{13}$CO
counterparts to $^{12}$CO structures}

\begin{figure*}
\begin{center}
\includegraphics[]{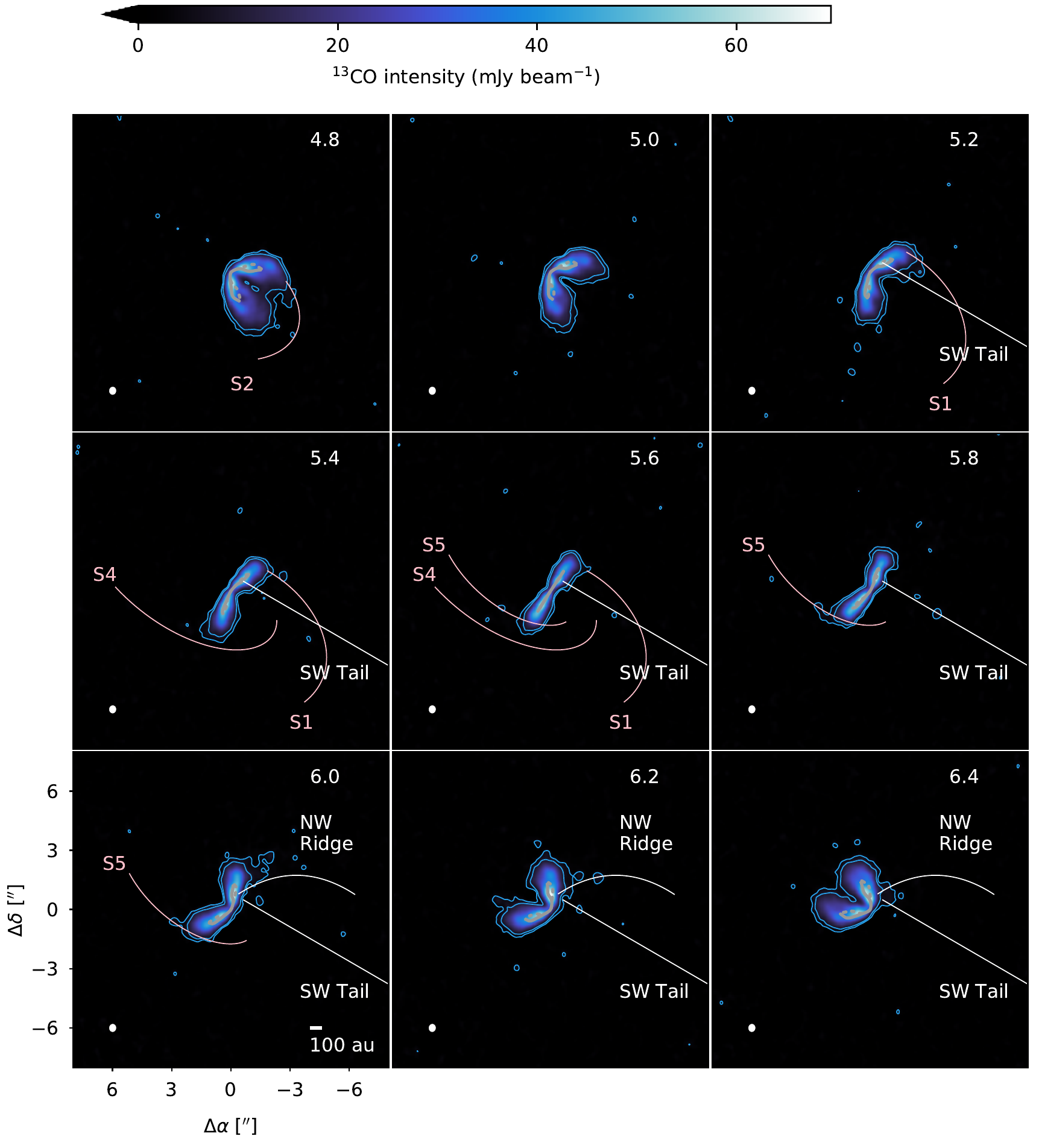}
\end{center}
\caption{$^{13}$CO $J=2-1$ channel maps annotated with the locations of $^{12}$CO features for which counterparts are tentatively identified in $^{13}$CO. The blue contours mark the 4 and 8$\sigma$ emission levels, where $\sigma=1$ mJy beam$^{-1}$. The Keplerian isovelocity contours are drawn in gray, with solid curves marking the front side of the disk and dashed curves marking the back side. The synthesized beam size, $0\farcs39\times0\farcs35$ (-4\fdg4), is represented with a white ellipse in the lower left corner of each panel. \label{fig:features13co}}
\end{figure*}
Figure \ref{fig:13COmommaps} shows the integrated intensity and intensity-weighted velocity maps for $^{13}$CO $J=2-1$ toward GM Aur. The $^{13}$CO integrated intensity map was made by summing over all channels from $-0.6$ to 11.8 km s$^{-1}$, with no mask applied. The intensity-weighted velocity map was computed over the same velocity range, with pixels below the $5\sigma$ level masked. In contrast to $^{12}$CO, $^{13}$CO does not exhibit strong deviations from axisymmetry or Keplerian motion in these maps. However, the channel maps exhibit some faint non-Keplerian structures that overlap with the inner portions of S1, S2, S4, S5, the northwest ridge, and the southwest tail (Figure \ref{fig:features13co}). No clear $^{13}$CO counterparts are identified for the northeast filamentary structures. 

None of the other lines observed as part of the MAPS Large Program exhibit obvious counterparts to the $^{12}$CO structures \citep{2021arXiv210906268O, 2021arXiv210906210L}. Most notably, other bright lines observed toward GM Aur include HCO$^+$ $J=1-0$, CS $J=2-1$, $^{13}$CO $J=1-0$, HCN $J=3-2$, and H$_2$CO $J_{K_aK_c} = 3_{03}-2_{02}$. The HCO$^+$ $J=3-2$ line was observed separately by \citet{2020ApJ...891...48H} at comparable spatial resolution and sensitivity, and also did not reveal any clear non-axisymmetric structures. The absence of counterparts identified in deep observations of these lines underlines the importance of $^{12}$CO as a probe of the complex surroundings of disks.

\subsection{Constraints on the mass of GM Aur's extended gas structures}
The $^{12}$CO and $^{13}$CO $J=2-1$ observations together can be used to place rough constraints on the mass of GM Aur's extended gas structures. In the optically thin, LTE limit, the molecular column density can be estimated as
\begin{equation}\label{eq:coldensity}
    N = \frac{Q(T)}{g}\exp{\left(\frac{E_u}{T}\right)}\frac{4\pi}{A_{ul}hc}\int I_\nu dv,
\end{equation}
where $Q$ is the partition function, $g$ is the upper state degeneracy, $E_u$ is the upper state energy, $T$ is the excitation temperature,  $A_{ul}$ is the Einstein $A$ coefficient, and $\int I_\nu dv$ is the integral of the intensity over the velocity axis \citep[e.g.,][]{1999ApJ...517..209G}.

Given the non-detection of most of the extended structures in lines other than $^{12}$CO $J=2-1$, neither the optical depth nor the gas temperature of the extended structures is well-constrained. As shown in Figure \ref{fig:COoverview}, the brightness temperatures of the extended structures range from $\sim10-20$ K, but it is not clear the extent to which these brightness temperature variations reflect column density versus gas temperature variations. While these would be relatively cold gas temperatures, they are not implausible. For the sake of estimating a lower bound in the gas mass from $^{12}$CO, we adopt the optically thin approximation.

\begin{figure}
\begin{center}
\includegraphics[]{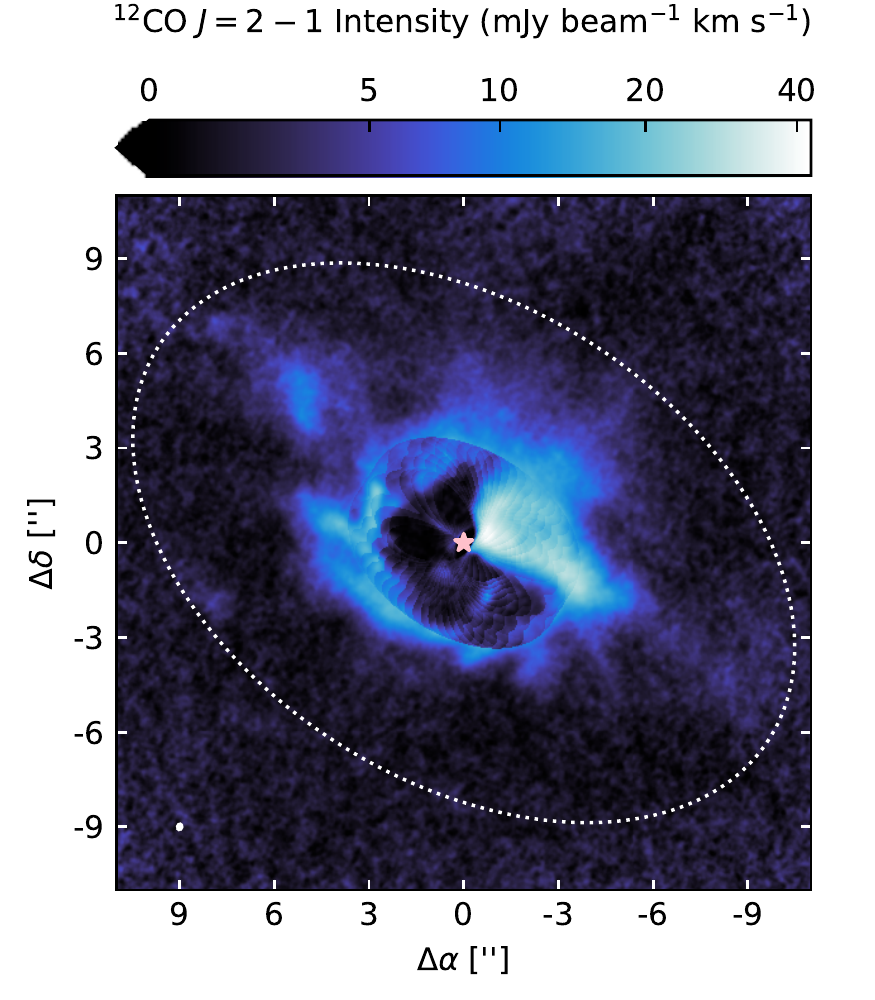}
\end{center}
\caption{An integrated intensity map of $^{12}$CO $J=2-1$ emission, with the Keplerian disk masked. Edge artifacts from the mask are visible in the central region of the image. An arcsinh color stretch is used to make faint structures more visible. The pink star marks the location of GM Aur. The dotted white ellipse denotes the area over which the flux of the extended structures is measured to estimate their gas mass. The synthesized beam is shown as an ellipse in the lower left corner. \label{fig:maskedmommap}}
\end{figure}

To exclude $^{12}$CO emission associated with the Keplerian disk from the mass estimate, we applied a Keplerian mask made with the \texttt{keplerian\_mask} package \citep{teague_kepmask} to the image cube. Details of the mask generation are provided in Section \ref{sec:keplerianmask}. An integrated intensity map was produced from the masked image cube, including velocities from 3.6 to 7.1 km s$^{-1}$ (i.e., channels where non-Keplerian structures are visible). Since spatial filtering led to some negative artifacts in the $^{12}$CO image cube, negative pixels were excluded from the integrated intensity map. The resulting map is shown in Figure \ref{fig:maskedmommap}.

The total flux of the extended structures was then extracted from an ellipse centered on GM Aur with a position angle of $57\fdg2$, semi-major axis of $11\farcs5$ (1830 au), and semi-minor axis of $7\farcs5''$ (1190 au), as drawn in Figure \ref{fig:maskedmommap}. Since the extent of the emission is comparable to the primary beam FWHM, it is possible that GM Aur's large-scale structures continue beyond ALMA's field of view. Even within ALMA's field of view,  the flux is likely underestimated due to spatial filtering. Thus, the flux measured from the ALMA observations should be treated as a lower bound. We find a total flux of 15 Jy km s$^{-1}$ and a mean integrated intensity of 4 mJy beam$^{-1}$ km s$^{-1}$ within the elliptical extraction region.

To convert the mean integrated intensity to a column density lower limit, we adopt an average gas temperature of 20 K, somewhat colder than the temperature that \citet{2021arXiv210906233Z}  derived for the CO emitting layer in the outer regions of GM Aur's Keplerian disk. The molecular constants for the $^{12}$CO $J=2-1$ transition, which are taken from the Cologne Database for Molecular Spectroscopy (CDMS), are $Q(20 \text{ K})= 7.6$, $g=5$, $E_u=16.6$ K, and $A_{ul}=6.91\times10^{-7}$ s$^{-1}$ \citep{2001AA...370L..49M, 2005JMoSt.742..215M}. Equation \ref{eq:coldensity} yields a $^{12}$CO column density lower limit of $7\times10^{14}$ cm$^{-2}$. Assuming an ISM-like $^{12}$CO:H$_2$ number density ratio of 10$^{-4}$ and that the gas is predominantly molecular hydrogen, we integrated the gas surface density over the area of the ellipse used to extract the $^{12}$CO flux and estimate that the gas mass is at least $2\times10^{-5}$ $M_\odot$.

Although the bulk of the extended $^{12}$CO structures are not detected in $^{13}$CO, $^{13}$CO can be used to set an upper bound on their gas mass. $I_\nu$ is approximated as a Gaussian line profile with a peak of 3 mJy beam$^{-1}$ (i.e., $3\times$ the rms) and an approximately thermal FWHM of 0.2 km s$^{-1}$. The molecular constants for the $^{13}$CO $J=2-1$ transition, which are taken from CDMS, are $Q(20 \text{ K})=15.8$, $g=10$, $E_u=15.9$ K and $A_{ul}=6.08\times10^{-7}$ s$^{-1}$ \citep{2001AA...370L..49M, 2005JMoSt.742..215M}. (Note that the upper state degeneracy differs from the LAMDA database value \citep{2005AA...432..369S} listed in \citet{2021arXiv210906268O} by a factor of 2 because the CDMS database accounts for hyperfine splitting in the partition function).  Equation \ref{eq:coldensity} yields a $^{13}$CO column density upper limit of $\sim6\times10^{13}$ cm$^{-2}$. Assuming that the $^{12}$C/$^{13}$C ratio has an ISM-like value of 69 \citep[e.g.,][]{1999RPPh...62..143W}, the $^{12}$CO:H$_2$ ratio is 10$^{-4}$, and that the gas is predominantly molecular hydrogen, we again integrate over the area of the ellipse used to extract the flux of the $^{12}$CO extended structures and estimate a nominal gas mass upper bound of 10$^{-4}$ $M_\odot$. This could still be an underestimate of the gas mass of the extended structures if the $^{12}$CO:H$_2$ ratio is substantially less than $10^{-4}$ due to freezeout, photodissociation, and/or chemical depletion. Models comparing GM Aur's CO column densities to the disk gas mass (as constrained by HD) suggest that even in the warm molecular layer, where neither freezeout nor photodissociation should significantly affect CO abundances, CO is still depleted by 1-2 orders of magnitude relative to ISM levels. Nevertheless, even with the uncertainty in the CO:H$_2$ ratio, the total mass contained in the extended structures appears to be considerably less than the gas mass of the Keplerian disk, which has been estimated to be $\sim0.02-0.2$ $M_\odot$ \citep[e.g.,][]{2016ApJ...831..167M, 2021arXiv210906228S}.

\section{Discussion}\label{sec:discussion}
\subsection{Possible origins of GM Aur's complex gas structures}
\subsubsection{Accretion of remnant envelope or cloud material}
In Section \ref{sec:diffuse}, we argued that GM Aur's extended gas kinematics are consistent with infalling motion. Morphologically, GM Aur's southwest tail and diffuse northern emission structures resemble scaled down versions of the infalling gas streams that have been detected in molecular emission toward some (younger) Class 0 and I sources \citep[e.g.,][]{2014ApJ...793....1Y, 2017AA...608A.134Y, 2020NatAs...4.1158P, 2020ApJ...904L...6A}. Hydrodynamical simulations also indicate that infalling material can induce the formation of spiral-like structures in or around disks \citep[e.g.,][]{2012MNRAS.427.1182S, 2015AA...582L...9L, 2015ApJ...805...15B, 2017ApJ...846....7K}. Thus, GM Aur's complex CO structures could result from ongoing accretion of material from the cloud or a remnant envelope. 

Notably, GM Aur underwent a ``burst'' in 2018, in which the stellar accretion rate changed by a factor of 3.5 within a couple of weeks \citep{2019ApJ...874..129R}. The accretion rate change coincided with a change in the disk's mid-infrared emission, leading \citet{2019ApJ...877L..34E} to hypothesize that the increased stellar accretion rate resulted from an increase in the inner disk's surface density. At earlier evolutionary stages, the high stellar accretion rates and powerful outbursts observed in FU Ori-like systems have often been attributed to instabilities triggered by a build-up of material in the disk due to envelope infall \citep[e.g.,][]{1996ARAA..34..207H, 2009ApJ...701..620Z, 2010ApJ...713.1143Z, 2014ApJ...795...61B}. Perhaps in an analogous fashion, infall from the large-scale gas structures around GM Aur may be responsible for its hypothesized inner disk surface density build-up and ``burst''-like behavior. Assessing the plausibility of such a link in the GM Aur system will require spatially resolved observations of other Class II systems exhibiting similar burst-like behavior. 

\begin{figure*}
\begin{center}
\includegraphics[]{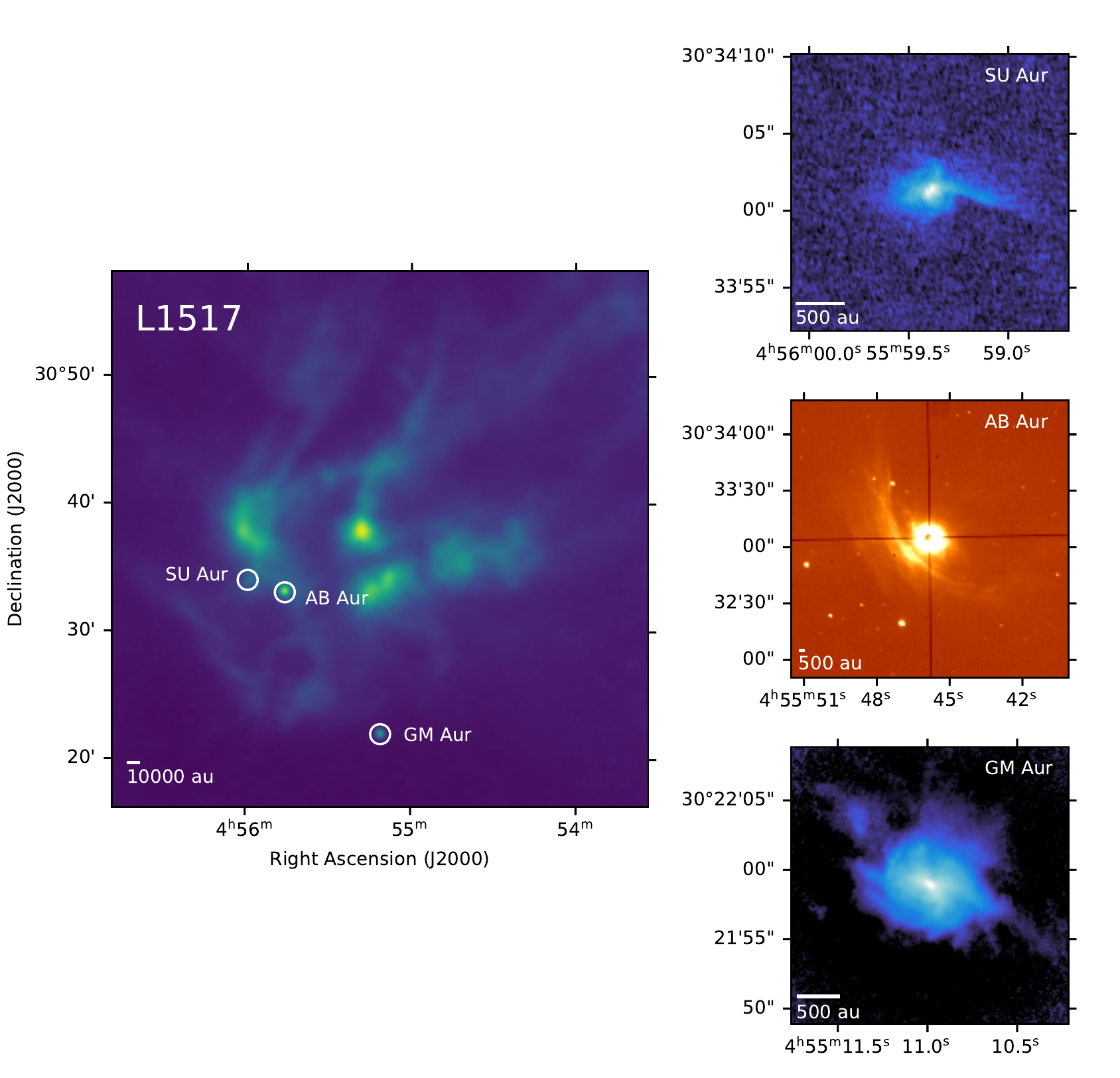}
\end{center}
\caption{Left: The positions of AB Aur, SU Aur, and GM Aur marked on a Herschel SPIRE \citep{2010AA...518L...1P, 2010AA...518L...3G} 500 $\mu$m map tracing dust emission in the L1517 cloud. The version of the map retrieved from the Herschel Science Archive is the Level 2.5 extended data processing product, a combination of ObsID 	
1342204843 and 	
1342204844 from the Herschel Gould Belt survey \citep{2010AA...518L.102A}. Top right: $^{12}$CO $J=3-2$ image of SU Aur \citep{2021ApJ...908L..25G}. Middle right: $R$-band image of AB Aur \citep{Kalasthesis, 1999ApJ...523L.151G}. Bottom right: $^{12}$CO $J=2-1$ image of GM Aur from this work.  \label{fig:L1517}}
\end{figure*}

A number of pre-main sequence stars have long been known to be surrounded by nebulosities detected in optical and infrared images \citep[e.g.,][]{1945ApJ...102..168J, 1960ApJS....4..337H}. The nature of the physical relationship between these stars and the nearby nebulosities, though, had not always been clear. However, increasingly powerful millimeter interferometers and high-contrast optical/near-infrared imagers have been able to resolve large-scale spirals, streams, and tails that can be traced down to disk scales \citep[e.g.,][]{1999ApJ...523L.151G, 2016SciA....2E0875L,  2020AA...633A..82G, 2020ApJ...898..140H}. It is interesting to note that GM Aur, AB Aur, and SU Aur, the three most extensively observed Class II systems in the L1517 cloud, all exhibit spiral structures in CO and/or scattered light, strong deviations from Keplerian kinematics, and connections (at least in projection) to large-scale arc or tail-like nebulosities detected in scattered light \citep[e.g.,][]{1999ApJ...523L.151G, 2003AJ....125.1467S,2012AA...547A..84T, 2015ApJ...806L..10D, 2019AJ....157..165A, 2020AA...637L...5B, 2021ApJ...908L..25G}. Figure \ref{fig:L1517} shows the locations of these systems in the L1517 cloud and images of the extended structures detected in association with their disks. Based on their kinematics and morphology, the AB Aur and SU Aur systems have often been cited as likely examples of disks undergoing late infall, where ``late'' generally refers to Class II disks more than 1 Myr old  \citep[e.g.,][]{1999ApJ...523L.151G,2012AA...547A..84T, 2019AJ....157..165A, 2021ApJ...908L..25G}. The similar properties of the extended gas structures of GM Aur, AB Aur, and SU Aur, in spite of their disparate stellar and millimeter continuum disk properties, lend further weight to the idea that the gas structures result from processes external to rather than within the disk. On a more speculative note, AB Aur and SU Aur's extended structures appear more prominent in CO and scattered light compared to those of GM Aur (see aforementioned references). In projection, AB Aur and SU Aur are positioned in higher-density regions of the L1517 cloud than GM Aur is (see Figure \ref{fig:L1517}). Observations of additional disks in L1517 could help to determine what relationship may exist between the large-scale distribution of material in the cloud and the local disk environment. 

Late infall has been hypothesized to provide access to a significant mass reservoir for planet formation, to misalign disks and therefore the orbits of planets that subsequently form, and to enhance stellar accretion rates, potentially leading to outbursts that heat up the disk \citep[e.g.,][]{2018AA...618L...3M, 2019AA...628A..20D, 2020AA...633A...3K}. Systematic surveys of disks in gas tracers will be fundamental for determining the prevalence of late infall and therefore the magnitude of its potential impact on planet demographics. While $^{12}$CO and scattered light observations can be used to establish whether the kinematics and morphology of the large-scale gas structures are indicative of late infall, they are not well-suited by themselves for assessing whether the mass or angular momentum transfer is sufficient to yield some of the dramatic effects predicted  by models. Thus, a critical step in assessing the role that late infall might play in planet formation is to search for suitable spatially resolved mass and temperature tracers via observations of other molecular lines. Potentially useful targets include CS or CN, since they have been shown sometimes to trace extended structures around embedded disks \citep[e.g.,][]{2021AA...645A.145G}. While no obvious large-scale structures are identified around GM Aur in CS $J=2-1$ emission by \citet{2021arXiv210906286L} or in CN $N=1-0$ by \citet{2021arXiv210906694B}, these structures might be recoverable in brighter transitions.

\subsubsection{Gravitational instability}
Gravitational instabilities (GI), potentially precipitated by infalling material, may drive spiral arm formation as well as the formation of tail-like structures through gas clumps ejected when the disk fragments \citep[e.g.,][]{1997Sci...276.1836B, 2001ApJ...553..174G,2005ApJ...633L.137V, 2008ApJ...681..375K, 2011MNRAS.413..423H,2020AA...635A.196V}. It is ambiguous from Toomre Q parameter estimates whether GM Aur is in fact unstable. From modeling HD emission, \citet{2016ApJ...831..167M} report that $Q\sim1.3$ at a radius of 300 au,  which would make the disk marginally gravitationally unstable. From modeling the same HD observations as well as new spatially resolved CO isotopologue observations from MAPS, \citet{2021arXiv210906228S} find that the disk may be gravitationally unstable between $\sim70-100$ au (where $Q<1.7$ is used as the metric), but stable at all other radii. This would suggest that GM Aur's spiral arms, which are located hundreds of au from the star, are not a consequence of GI. However, \citet{2021arXiv210906228S} note that their surface densities beyond 100 au may be underestimated (and therefore the Toomre Q parameter overestimated) because HD emission comes from warmer gas. Using multi-frequency millimeter continuum observations and assuming a gas to dust ratio of 100, \citet{2021arXiv210906433S} find that the disk is gravitationally stable between 25 and 125 au, with Q values that are systematically higher than those estimated by \citet{2021arXiv210906228S}. Because the millimeter continuum emission is compact compared to the gas, the Q parameter estimates from \citet{2021arXiv210906433S} are not directly relevant to assessing the disk stability at the location of the spiral arms. However, the discrepant values that have been estimated for GM Aur highlight the challenges of assessing disk stability. 

In principle, disk kinematics offer an independent avenue to assess whether a disk is unstable. \citet{2020ApJ...904..148H} use hydrodynamical simulations to demonstrate that GI should induce a characteristic non-Keplerian ``wiggle'' in CO channel maps. Given the overall kinematic and geometric complexity of GM Aur's CO emission, it is not obvious whether such a ``wiggle'' is present. We note that in any case, disk instabilities alone would not completely account for GM Aur's morphology, particularly the diffuse northern structures. 

\subsubsection{Other explanations}

Simulations have shown that close stellar encounters can create spiral and tail structures in disks \citep[e.g.,][]{1993MNRAS.261..190C, 2003ApJ...592..986P, 2019MNRAS.483.4114C, 2020AA...635A.196V}. Gas structures reminiscent of the GM Aur system have also been detected in several systems with stars separated by projected distances of a few hundred au, including RW Aur A and B \citep{2018ApJ...859..150R}, AS 205 N and S \citep{2018ApJ...869L..44K}, and UX Tau A and C \citep{2020ApJ...896..132Z}. 

Unlike these other systems, though, there are no strong candidates for objects that have undergone close encounters with GM Aur. Simulations indicate that following an encounter, spiral arms should only persist in the disk for several thousand years \citep[e.g.,][]{2019MNRAS.483.4114C}. The one-dimensional velocity dispersion of the Taurus molecular cloud complex has been estimated to be $\sim3$ km s$^{-1}$ \citep{2018ApJ...859...33G}. Supposing that GM Aur experienced an encounter 5000 years ago with a star moving away at a relatively large velocity of 10 km s$^{-1}$, their present-day separation would be $\sim11,000$ au. Of the Taurus members identified in \citet{2018AJ....156..271L} and \citet{2019AJ....158...54E} from the Gaia mission \citep{2016AA...595A...2G, 2018AA...616A...1G} and follow-up observations, the closest source to GM Aur is 2MASS J04552333+3027366, a brown dwarf located 6.2 arcmin (59,300 au) northeast of GM Aur. Another source with a similar parallax, Gaia DR2 156913778302440704, is located closer in projection, $45''$ (7,200 au) northeast of GM Aur \citep{2016AA...595A...2G, 2018AA...616A...1G}. However, Gaia DR2 156913778302440704 is not classified as a member of Taurus in \citet{2018AJ....156..271L} because its Gaia-measured proper motion ($\mu_\alpha=10.5\pm0.2$ mas yr$^{-1}$, $\mu_\delta=-32.8\pm0.1$ mas yr$^{-1}$) differs significantly from the median proper motion measured for the L1517 region ($\mu_\alpha=4.7\pm0.8$ mas yr$^{-1}$, $\mu_\delta=-24.5\pm1.7$ mas yr$^{-1}$). \citet{2018AJ....156..271L} estimated that in regions where $A_J<1$, their Taurus catalog is complete for spectral types earlier than M6-M7, which corresponds to a magnitude of $G\sim19$. This limit is valid in the vicinity of GM Aur, for which $A_J=0.14$ \citep{2017AJ....153...46L}. \citet{2018AJ....156..271L} estimated that at an age of several million years (i.e., roughly that of the Taurus star-forming region), spectral types of M6-M7 correspond to stellar masses of roughly 0.1 $M_\odot$. Meanwhile, \citet{2018AJ....156..259Z} determined that Gaia can recover nearly all secondary companions with $G<21$ and separations greater than $3''$. In addition, the SEEDS high-contrast imaging survey, with a field of view of $\sim400$ au, rules out companions more massive than $\sim2.5$ $M_{\text{Jup}}$ beyond 50 au from GM Aur \citep{2017AJ....153..106U}. Together, these studies appear to rule out the presence of stellar mass objects in GM Aur's vicinity. A more qualitative argument against GM Aur having undergone a recent stellar encounter is that stellar encounters are expected to lead to tidal truncation of disks \citep[e.g.,][]{2014AA...565A.130B}, but the GM Aur disk has one of the largest known radial extents among Class II disks in both molecular and millimeter continuum emission \citep[e.g.,][]{1993Icar..106....2K, 2017ApJ...845...44T}. 

Stellar or planetary companions can also excite spiral density waves \citep[e.g.,][]{1986ApJ...307..395L, 2002ApJ...565.1257T}. Stellar companions appear unlikely to be the cause of GM Aur's spiral structures for the same reasons that stellar encounters are not deemed likely to be responsible. Meanwhile, planetary mass companions are expected to result in small and localized perturbations to the disk gas kinematics rather than the large-scale non-Keplerian motion observed toward GM Aur \citep[e.g.,][]{2018ApJ...860L..13P, 2018MNRAS.480L..12P, 2019Natur.574..378T}. At any rate, a planetary mass companion would not account for either the southwest tail or diffuse northern emission. 

\subsection{Exploring the chemical implications of GM Aur's complex gas structures}

Given that GM Aur was observed as part of the MAPS disk chemistry survey, it is worthwhile to consider how the processes that might be responsible for GM Aur's extended CO structures could also influence the disk chemistry, and how best to test these ideas. Chemically speaking, the outer disk of GM Aur probed at millimeter wavelengths is notable for both its faint C$_2$H and bright H$_2$CO emission (adjusted for distance) relative to similarly large, massive disks \citep{2016AA...592A.124G, 2019ApJ...876...25B, 2020ApJ...890..142P, 2021arXiv210906268O, 2021arXiv210906391G}

At first glance, GM Aur's properties suggest that it should be hospitable to C$_2$H production. The factors thought to be most critical for C$_2$H production in disks are high volatile C/O ratios and strong UV radiation \citep[e.g.,][]{2016ApJ...831..101B, 2021arXiv210906221B}. While the C/O ratio has not been measured directly for GM Aur, its gas-phase CO has been estimated to be depleted up to two orders of magnitude \citep[e.g.,][]{2016ApJ...831..167M, 2021arXiv210906233Z,2021arXiv210906228S}. High volatile CO depletion is often thought signify elevated C/O ratios \citep[e.g.,][]{2016AA...592A..83K, 2016ApJ...831..101B, 2019AA...631A..69M}. Meanwhile, GM Aur's UV radiation field (as quantified by $L_{Ly\alpha}/L_{\text{UV}}$ and $L_{\text{FUV}}/L_{\text{UV}}$) is comparable to other Class II disks known to have bright C$_2$H emission \citep[e.g.,][]{2020AJ....159..168A}. One possible explanation for GM Aur's more modest C$_2$H emission is that if the disk is indeed still accreting cloud or envelope material, then enough oxygen from either CO or photodesorbing water ice might be introduced into the surface layers to reduce hydrocarbon production without resetting the chemistry in the deeper layers containing the bulk of the disk mass. Astrochemical modeling of infalling material onto a Class II disk will be useful for quantifying whether this mechanism accounts for the observed hydrocarbon behavior. 

Although a large disk mass could be a straightforward explanation for GM Aur's bright H$_2$CO emission, it is still notable insofar as some of GM Aur's other molecular lines are faint compared to other sources with comparable or lower masses \citep{2020ApJ...890..142P, 2021arXiv210906268O, 2021arXiv210906391G}. While the number of measurements is small, observations of Class I disks suggest that their H$_2$CO column densities tend to be higher than in Class II disks \citep{2020ApJ...901..166V, 2021AA...645A.145G}. If GM Aur is accreting primordial gas, some of its chemical characteristics may bear closer resemblance to a young, embedded disk than to other Class II disks. A possible reason for GM Aur's bright H$_2$CO emission could be shock heating induced by infall and/or disk instabilities, which in turn may lead to enhanced H$_2$CO desorption from grains as well as increased gas-phase production \citep[e.g.][]{2011MNRAS.417.2950I, 2019MNRAS.483.1266E}. An excitation temperature measurement of H$_2$CO in this disk would be useful to ascertain whether shock heating is indeed occurring. 

While GM Aur has been a fairly popular target for disk chemistry observations, interferometric studies of molecular emission have otherwise preferentially targeted sources exhibiting relatively clean signatures of Keplerian rotation in order to avoid confusion with cloud material \citep[e.g.][]{2007AA...467..163P, 2008ApJ...681.1396Q,2010ApJ...720..480O,2016ApJ...831..101B}. By definition, choosing targets based on their Keplerian signatures has limited the extent to which Class II disks undergoing infall have been characterized. A broader sample needs to be observed to investigate whether there are indeed systematic chemical differences between sources showing evidence of interacting with their surroundings versus those that are more isolated.

\section{Summary and Conclusions}\label{sec:conclusions}

As part of the MAPS Large Program, we observed CO toward the GM Aur disk at high spatial resolution and sensitivity. Our major findings are as follows: 

\begin{enumerate}
    \item In addition to a Keplerian disk visible up to a radius of $\sim550$ au, $^{12}$CO emission reveals spiral arms out to a radius of $\sim$1200 au, a tail extending up to $\sim1800$ au southwest of GM Aur, and diffuse northern structures up to $\sim1900$ au from the star. These large-scale non-axisymmetric gas structures contrast sharply with the nearly axisymmetric, multi-ringed millimeter continuum, which is detected only up to $\sim250$ au from the star.
    \item Portions of the diffuse northern structures and southwest tail have counterparts in earlier HST scattered light observations tracing the small grain distribution. 
    \item Based on the kinematics and morphology of GM Aur's extended gas structures, we hypothesize that the GM Aur disk is experiencing late infall, either from a remnant protostellar envelope or from surrounding cloud material.  
    \item The extended $^{12}$CO structures are only tentatively detected in $^{13}$CO. The absence of strong extended emission in other molecular tracers underscores the utility of observing bright $^{12}$CO lines to probe interactions between disks and their surroundings. 
    \item Given GM Aur's relatively weak C$_2$H and strong H$_2$CO emission in comparison with other massive Class II disks that have been observed, we speculate that late infall could serve as a ``fountain of youth'' that partially resets disk chemistry. Molecular line observations of other systems exhibiting signatures of late infall will be necessary in order to establish if and how environmental interactions affect disk and protoplanet compositions.

\end{enumerate}

While recent ALMA surveys have taken advantage of the array's unparalleled spatial resolution and sensitivity to probe the inner regions of protoplanetary disks,  serendipitous detections of extended gas structures such as spirals and tails have shown that much is left to be learned about the structure and evolution of disks on larger scales as well. Ascertaining how disks interact with their surroundings through deep, systematic $^{12}$CO observations will be key for determining how protoplanets ultimately acquire their starting material. 

\vspace{5mm}
\facilities{ALMA, HST (WFPC2), Herschel}

\software{\texttt{analysisUtils} (\url{https://casaguides.nrao.edu/index.php/Analysis\_Utilities}), \texttt{AstroPy} \citep{2013AA...558A..33A}, \texttt{CASA} \citep{2007ASPC..376..127M}, \texttt{gofish} \citep{2019JOSS....4.1632T}, \texttt{keplerian\_mask} \citep{teague_kepmask},  \texttt{matplotlib} \citep{Hunter:2007}, \texttt{numba}  \citep{10.1145/2833157.2833162}, \texttt{NumPy} \citep{2020NumPy-Array}, \texttt{scikit-image} \citep{scikit-image}, \texttt{SciPy} \citep{2020SciPy-NMeth}}

\acknowledgments
This paper makes use of ALMA data \dataset[ADS/JAO.ALMA\#2018.1.01055.L]{https://almascience.nrao.edu/aq/?project\_code=2018.1.01055.L}, \\ \dataset[ADS/JAO.ALMA\#2017.1.01151.S]{https://almascience.nrao.edu/aq/?project\_code=2017.1.01151.S}, and \\ \dataset[ADS/JAO.ALMA\#2018.1.01230.S]{https://almascience.nrao.edu/aq/?project\_code=2018.1.01230.S}. ALMA is a partnership of ESO (representing its member states), NSF (USA) and NINS (Japan), together with NRC (Canada) and NSC and ASIAA (Taiwan), in cooperation with the Republic of Chile. The Joint ALMA Observatory is operated by ESO, AUI/NRAO and NAOJ. The National Radio Astronomy Observatory is a facility of the National Science Foundation operated under cooperative agreement by Associated Universities, Inc. This research is based on observations made with the NASA/ESA Hubble Space Telescope obtained from the Space Telescope Science Institute, which is operated by the Association of Universities for Research in Astronomy, Inc., under NASA contract NAS 5-26555. These observations are associated with program(s) HST-GO-6223. Herschel is an ESA space observatory with science instruments provided by European-led Principal Investigator consortia and with important participation from NASA. SPIRE has been developed by a consortium of institutes led by Cardiff University (UK) and including Univ. Lethbridge (Canada); NAOC (China); CEA, LAM (France); IFSI, Univ. Padua (Italy); IAC (Spain); Stockholm Observatory (Sweden); Imperial College London, RAL, UCL-MSSL, UKATC, Univ. Sussex (UK); and Caltech, JPL, NHSC, Univ. Colorado (USA). This development has been supported by national funding agencies: CSA (Canada); NAOC (China); CEA, CNES, CNRS (France); ASI (Italy); MCINN (Spain); SNSB (Sweden); STFC, UKSA (UK); and NASA (USA).  This research has made use of data from the Herschel Gould Belt survey (HGBS) project (http://gouldbelt-herschel.cea.fr). The HGBS is a Herschel Key Programme jointly carried out by SPIRE Specialist Astronomy Group 3 (SAG 3), scientists of several institutes in the PACS Consortium (CEA Saclay, INAF-IFSI Rome and INAF-Arcetri, KU Leuven, MPIA Heidelberg), and scientists of the Herschel Science Center (HSC). This research has made use of NASA's Astrophysics Data System and the VizieR catalogue access tool, CDS, Strasbourg, France. 

Support for J.H. was provided by NASA through the NASA Hubble Fellowship grant \#HST-HF2-51460.001-A awarded by the Space Telescope Science Institute, which is operated by the Association of Universities for Research in Astronomy, Inc., for NASA, under contract
NAS5-26555. J.H. and S.M.A. acknowledge funding support from the National Aeronautics and Space Administration under Grant No. 17-XRP17 2-0012 issued through the Exoplanets Research Program. E.A.B. and A.D.B. acknowledge support from NSF AAG Grant \#1907653. K.I.\"O. acknowledges support from the Simons Foundation (SCOL \#321183) and an NSF AAG Grant (\#1907653). R.T. and F.L. acknowledge support from the Smithsonian Institution as Submillimeter Array (SMA) Fellows. C.J.L. acknowledges funding from the National Science Foundation Graduate Research Fellowship under Grant No. DGE1745303. P.K. was supported by NNX15AD95G, part of NASA's Nexus for Exoplanet System Science (NExSS) research coordination network sponsored by NASA's Science Mission Directorate. Y.A. acknowledges support by NAOJ ALMA Scientific Research Grant Code 2019-13B and Grant-in-Aid for Scientific Research (S) 18H05222. J.B. acknowledges support by NASA through the NASA Hubble Fellowship grant \#HST-HF2-51427.001-A awarded  by  the  Space  Telescope  Science  Institute,  which  is  operated  by  the  Association  of  Universities  for  Research  in  Astronomy, Incorporated, under NASA contract NAS5-26555. J.B.B. acknowledges support from NASA through the NASA Hubble Fellowship grant \#HST-HF2-51429.001-A, awarded by the Space Telescope Science Institute, which is operated by the Association of Universities for Research in Astronomy, Inc., for NASA, under contract NAS5-26555. A.S.B acknowledges the studentship funded by the Science and Technology Facilities Council of the United Kingdom (STFC). J.K.C. acknowledges support from the National Science Foundation Graduate Research Fellowship under Grant No. DGE 1256260 and the National Aeronautics and Space Administration FINESST grant, under Grant no. 80NSSC19K1534. G.C. is supported by NAOJ ALMA Scientific Research Grant Code. 2019-13B. L.I.C. gratefully acknowledges support from the David and Lucille Packard Foundation and Johnson \& Johnson's WiSTEM2D Program. I.C. was supported by NASA through the NASA Hubble Fellowship grant HST-HF2-51405.001-A awarded by the Space Telescope Science Institute, which is operated by the Association of Universities for Research in Astronomy, Inc., for NASA, under contract NAS5-26555. J.D.I. acknowledges support from the Science and Technology Facilities Council of the United Kingdom (STFC) under ST/T000287/1. R.L.G. acknowledges support from a CNES fellowship grant. F.M. acknowledges support from ANR of France under contract ANR-16-CE31-0013 (Planet-Forming-Disks) and ANR-15-IDEX-02 (through CDP ``Origins of Life''). H.N. acknowledges support by NAOJ ALMA Scientific Research Grant Code 2018-10B and Grant-in-Aid for Scientific Research 18H05441. K.R.S. acknowledges the support of NASA through the NASA Hubble Fellowship grant \#HST-HF2-51419.001 awarded by the Space Telescope Science Institute, which is operated by the Association of Universities for Research in Astronomy, Inc., for NASA, under contract NAS5-26555. T.T. is supported by JSPS KAKENHI Grant Numbers JP17K14244 and JP20K04017. M.L.R.H. acknowledges support from the Michigan Society of Fellows. Y.Y. is supported by IGPEES, WINGS Program, the University of Tokyo. C.W. acknowledges financial support from the University of Leeds, STFC and UKRI (grant numbers ST/R000549/1, ST/T000287/1, MR/T040726/1). K.Z. acknowledges the support of the Office of the Vice Chancellor for Research and Graduate Education at the University of Wisconsin-Madison with funding from the Wisconsin Alumni Research Foundation, and the support of NASA through Hubble Fellowship grant HST-HF2-51401.001. awarded by the Space Telescope Science Institute, which is operated by the Association of Universities for Research in Astronomy, Inc., for NASA, under contract NAS5-26555. We thank Robert Gutermuth, Glenn Schneider, Jonathan Williams, Alvaro Hacar, and Megan Ansdell for useful discussions, as well as the anonymous referee for comments improving the manuscript. 

\appendix
\section{$^{12}$CO channel maps}\label{sec:chanmaps}
Figure \ref{fig:COfullchanmaps} presents $^{12}$CO $J=2-1$ channel maps toward GM Aur without annotations. 
\begin{figure*}
\begin{center}
\includegraphics{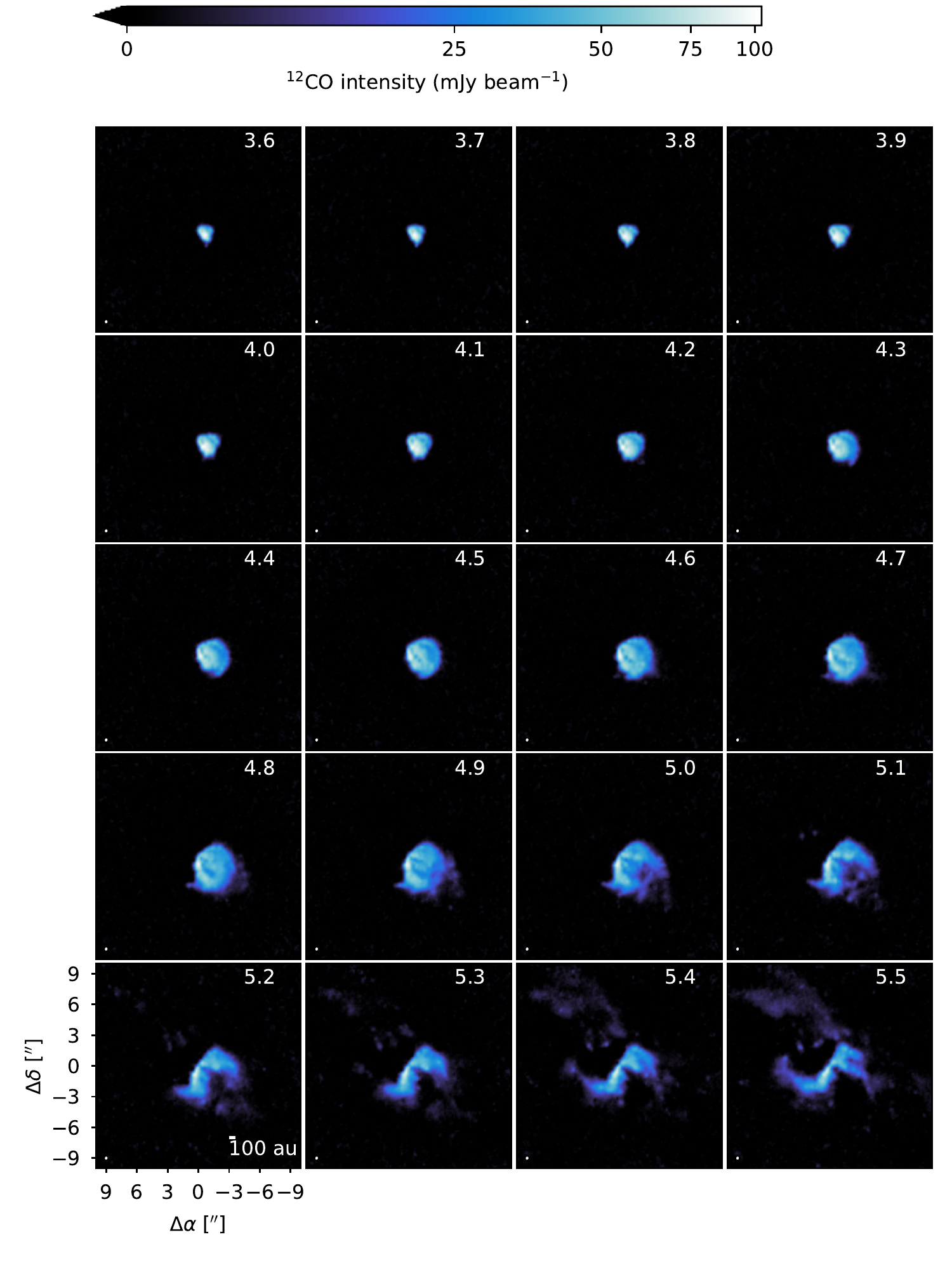}
\end{center}
\caption{Channel maps of the $^{12}$CO $J=2-1$ emission toward the GM Aur protoplanetary disk. The synthesized beam is shown in the lower left corner and the LSRK velocity (km s$^{-1}$) is labeled in the upper right corner of every panel. Offsets from the disk center (arcseconds) are marked on the axes in the lower left corner of the figure. An arcsinh color stretch is used to highlight faint emission. North is up and east is to the left. \label{fig:COfullchanmaps}}
\end{figure*}

\begin{figure*}
\begin{center}
\ContinuedFloat
\includegraphics{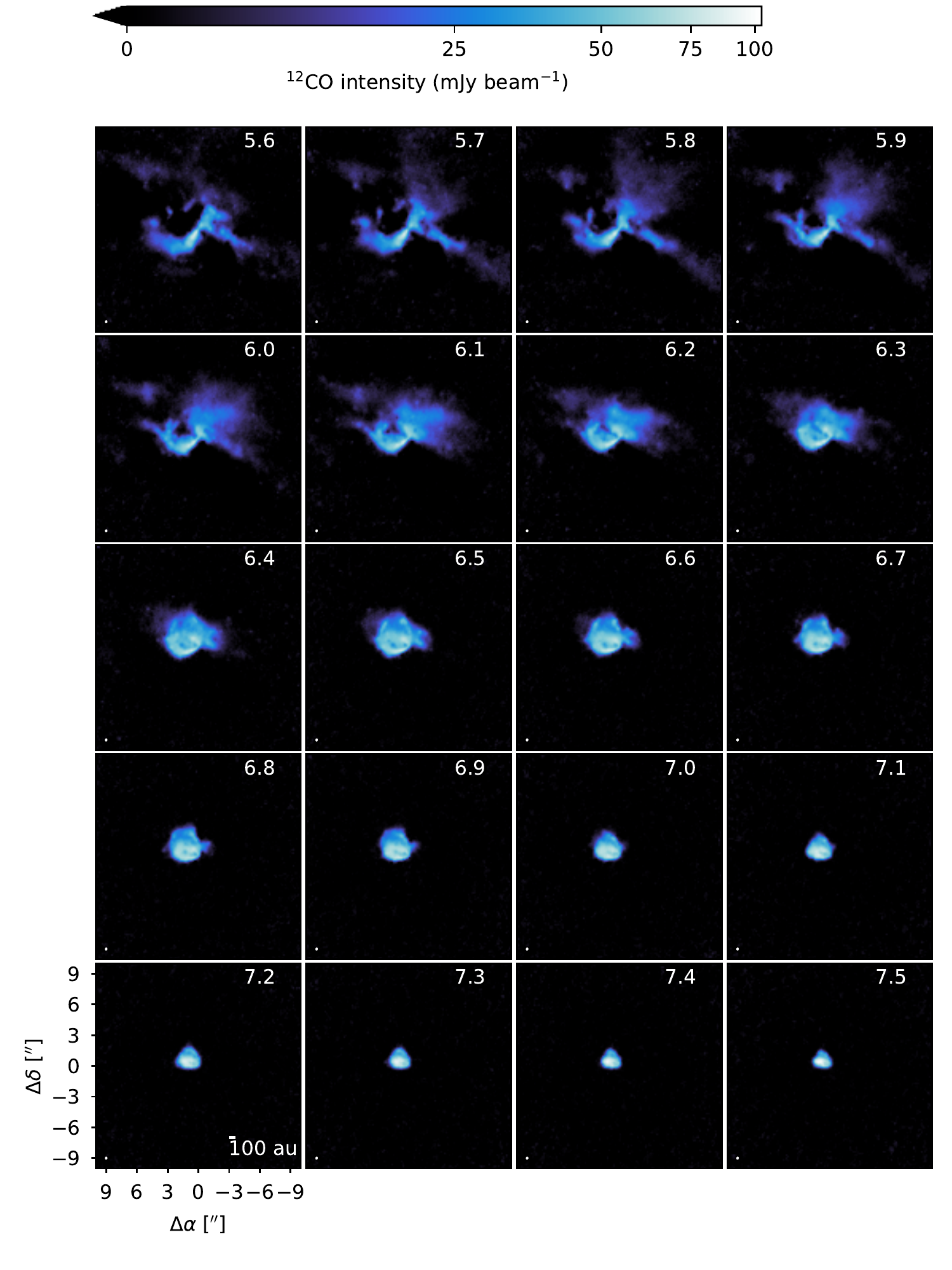}
\end{center}
\caption{Continued}
\end{figure*}

\begin{figure*}
\begin{center}
\includegraphics{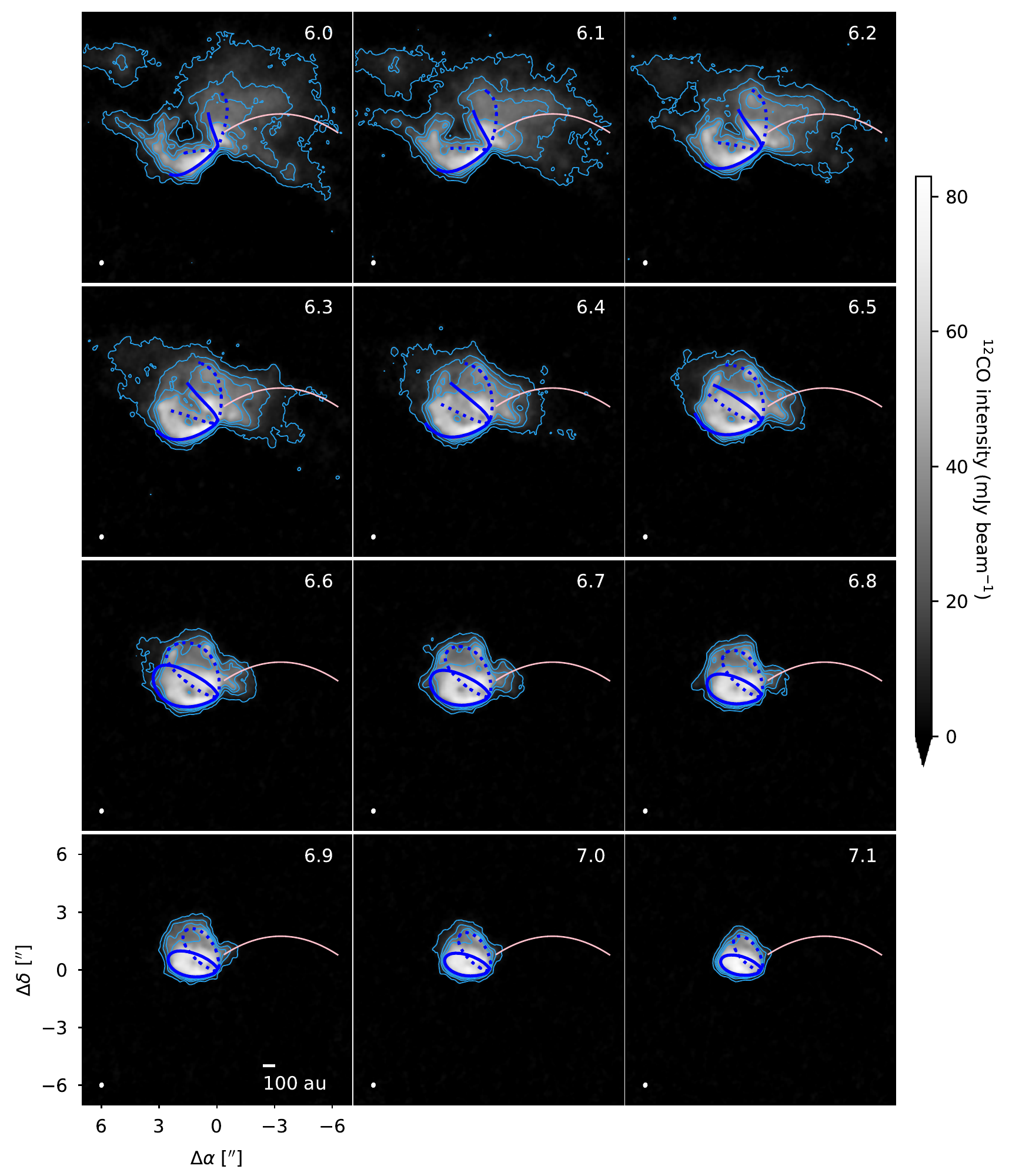}
\end{center}
\caption{Channel maps of the $^{12}$CO $J=2-1$ emission toward GM Aur with the best-fit curve to the northwest ridge overlaid in pink. The light blue contours show the 5, 15, 25, and 35$\sigma$ emission levels, where $\sigma= 1$ mJy beam$^{-1}$. The dark blue curves show the isovelocity contours of the Keplerian disk, with solid curves denoting the front of the disk and dashed curves denoting the back side. \label{fig:northwestchanmaps}}
\end{figure*}

Figure \ref{fig:northwestchanmaps} shows the best-fit curve to the northwest ridge overlaid on the $^{12}$CO channel maps. 

\section{Computing isovelocity contours for GM Aur's Keplerian disk }\label{sec:isovelocities}

To compute the isovelocity contours approximately corresponding to the portion of $^{12}$CO $J=2-1$ emission tracing GM Aur's Keplerian disk, we adopt the following emitting height $z$ (au) measured by \citet{2021arXiv210906217L}:

\begin{equation}
    z(r) = \text{61.2 au}\left( \frac{r}{\text{159 au}}\right)^{1.066} \exp{\left[-\left( \frac{r}{\text{599.0  au}}\right)^{4.988}\right]}, 
\end{equation}
where $r$ is the disk radius (au) in cylindrical coordinates. The azimuthal velocity of the gas $v_\varphi$ in the frame of the disk is calculated with the expression 

\begin{equation}\label{eq:vphi}
\frac{v_\varphi^2}{r} = \frac{GM_\ast r}{(r^2+z^2)^\frac{3}{2}}+\frac{1}{\rho(r,z)}\frac{\partial P(r,z)}{\partial r},
\end{equation}
where $\rho$ is the gas density and $P$ is the gas pressure. The two terms on the right-hand side account for the effects of stellar gravity and the gas pressure gradient, respectively. A stellar mass of $M_\ast = 1.1$ $M_\odot$ is adopted \citep{2018ApJ...865...37M}. The gas pressure gradient term is often dropped in velocity calculations because purely Keplerian rotation can usually reproduce disk molecular emission well \citep[e.g.,][]{1994AA...286..149D, 2000ApJ...545.1034S, 2015ApJ...806..154C}. However, deviations from Keplerian motion due to the pressure gradient become more prominent at large disk radii and heights, so the effect of the pressure gradient should be considered in the analysis of well-resolved $^{12}$CO observations of large disks \citep[e.g.,][]{2013ApJ...774...16R, 2020AA...633A.137D}. For brevity, even though the disk velocities are slightly sub-Keplerian, we refer to the $^{12}$CO isovelocity contours calculated from Equation \ref{eq:vphi} as the isovelocity contours of the Keplerian disk to make a distinction from the gas structures tracing the spiral arms, southwest tail, and diffuse northern structures. Equation \ref{eq:vphi} neglects self-gravity, which may not be accurate for the GM Aur disk because some estimates place its mass at or close to the gravitationally unstable limit \citep[e.g.,][]{2016ApJ...831..167M, 2021arXiv210906228S}. While we can produce isovelocity contours that match reasonably well with GM Aur's emission without considering self-gravity, it should be kept in mind that the isovelocity contour calculations are intended to provide a visual aid rather than a model that captures the full physical complexity of the system.  

The pressure is given by $P(r,z) = \rho(r,z) c_s^2$, where $c_s = \sqrt{\frac{k_B T(r)}{\mu m_H}}$ is the isothermal sound speed. We set $\mu=2.37$. Based on the GM Aur $^{12}$CO $J=2-1$ brightness temperatures measured in \citet{2021arXiv210906217L}, the temperature structure is approximated as 
\begin{equation}
T(r) = \text{52 K}\left(\frac{r}{\text{100 au}}\right)^{-0.608}.
\end{equation} 
The gas density profile is taken from the GM Aur model presented in \citet{2021arXiv210906233Z}.  The disk gas surface density is 
\begin{equation}
    \Sigma_g (r) = \text{9.4 g cm}^{-2} \left(\frac{r}{\text{176 au}} \right)^{-1} \exp{\left[-\left(\frac{r}{\text{176 au}} \right)\right]}.
\end{equation}
GM Aur's gas density is approximated as
\begin{equation}
\rho(r,z) = \frac{\Sigma_g(r)}{\sqrt{2\pi} H(r)} \exp{\left( -\frac{z^2}{2H(r)^2}\right)},
\end{equation}
where 
\begin{equation}
H(r) = 7.5 \text{ au} \left( \frac{r}{100 \text{ au}}\right)^{1.35}
\end{equation}
is the gas pressure scale height. 

The azimuthal velocity is related to the line-of-sight velocity $v_{\text{los}}$ by 
\begin{equation}
v_{\text{los}} = v_{\text{sys}} + v_\varphi \cos\varphi \sin i,
\end{equation}
where $i=53\fdg21$ is the disk inclination and $v_{\text{sys}}=5.61$ km s$^{-1}$ is the systemic velocity \citep{2020ApJ...891...48H}. Using the \texttt{gofish} Python package \citep{2019JOSS....4.1632T}, the position of each pixel in the channel maps in the frame of the observer is mapped to the $r,\varphi,z$ coordinates of the upper and lower surfaces of the $^{12}$CO emission in the frame of the disk in order to compute the line of sight velocity at each position. 

The $^{13}$CO $J=2-1$ isovelocity contours are computed similarly based on the emission height and brightness temperature measurements from \citet{2021arXiv210906217L}: 
\begin{equation}
    z(r) = \text{18.0 au}\left( \frac{r}{\text{159 au}}\right)^{4.539} \exp{\left[-\left( \frac{r}{\text{237.9 au}}\right)^{4.989}\right]}
\end{equation}
and 
\begin{equation}
T(r) = \text{22 K}\left(\frac{r}{\text{100 au}}\right)^{-0.260},
\end{equation}
respectively. 

\section{Constructing a Keplerian mask for GM Aur's $^{12}$CO emission }\label{sec:keplerianmask}
\begin{figure*}
\begin{center}
\includegraphics{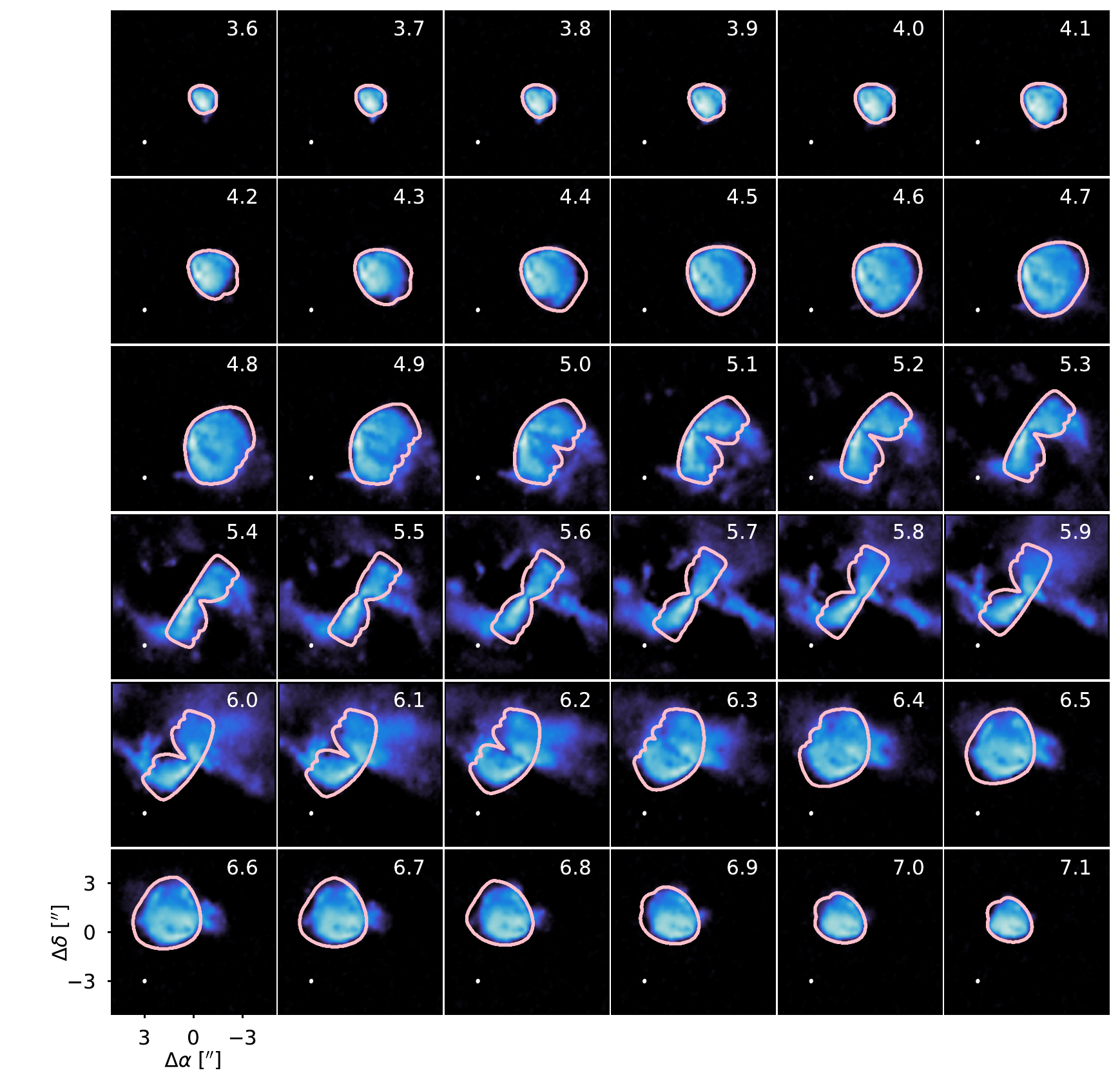}
\end{center}
\caption{The $^{12}$CO Keplerian mask (outlined in pink) used to produce the integrated intensity map shown in Figure \ref{fig:maskedmommap}, overlaid onto the $^{12}$CO channel maps. The LSRK velocity (km s$^{-1}$) is noted in the upper right corner of each panel and the synthesized beam is shown in the lower left corner. \label{fig:keplerianmask}}
\end{figure*}

Because GM Aur's extended $^{12}$CO structures overlap in projection with the Keplerian disk but have different line-of-sight velocities, we use a Keplerian mask to isolate emission originating from the extended structures. The \texttt{keplerian\_mask} Python package \citep{teague_kepmask} can calculate boolean masks based on the expected emitting region of a Keplerian disk for a given stellar mass, disk inclination and position angle, emitting surface height $z(r)$, and disk radius. We used the same stellar mass, disk orientation, and $z(r)$ used to draw the isovelocity contours described in Section \ref{sec:isovelocities}. The maximum radius was set to $r=550$ au based on visual inspection. The code also assumes that the intrinsic linewidth has the form
\begin{equation}
\Delta V(r) = \Delta V_0 \times \left( \frac{r}{1\arcsec} \right)^{\Delta V_q}.
\end{equation}
We used the code's default values of $\Delta V_0 = 300$ m s$^{-1}$ and $\Delta V_q = -0.5$ because they yielded a mask that reasonably encompassed the emission from the Keplerian disk, but the adopted linewidth profile should not necessarily be taken as a close representation of the true disk conditions. Finally, the mask was convolved with a two-dimensional Gaussian with the same size as the synthesized beam in order to account for emission broadening due to resolution limits. Figure \ref{fig:keplerianmask} shows the mask overlaid on the $^{12}$CO channel maps.

\bibliography{../allreferences}{}
\bibliographystyle{aasjournal}

\end{document}